\documentclass[aps,preprint,nofootinbib,eqsecnum]{revtex4}
\usepackage{amsmath}
\usepackage{amsfonts}
\usepackage{amssymb}
\usepackage{graphicx}

\input{tcilatex}

\begin{document}

\title{{\Large Hidden 12-dimensional Structures in }\\
AdS$_{5}\times$S$^{5}$ and M$^4\times$R$^6$ Supergravities}
\author{\textbf{{\large {Itzhak Bars}}}{\thanks{
Research partially supported by the US Department of Energy under grant
number DE-FG03-84ER40168.}}}

\begin{abstract}
It is shown that AdS$_{5}\times $S$^{5}$ supergravity has hitherto unnoticed
supersymmetric properties that are related to a hidden 12-dimensional
structure. The totality of the AdS$_{5}\times $S$^{5}$ supergravity
Kaluza-Klein towers is given by a single superfield that describes the
quantum states of a 12-dimensional supersymmetric particle. The particle has
super phase space $\left( X^{M},P^{M},\Theta \right),$ with $\left(
10,2\right) $ signature and 32 fermions. The worldline action is constructed
as a generalization of the supersymmetric particle action in Two-Time
Physics. SU$\left( 2,2|4\right) $ is a linearly realized global
supersymmetry of the 2T action. The action is invariant under the gauge
symmetries Sp $\left( 2,R\right) ,$ SO$\left( 4,2\right) ,$ SO$\left(
6\right) ,$ and fermionic kappa. These gauge symmetries insure unitarity and
causality while allowing the reduction of the 12-dimensional super phase
space to the correct super phase space for AdS$_{5}\times $S$^{5}$ or $%
M^{4}\times R^{6}$ with 16 fermions and one time, or other dually related
one time spaces. One of the predictions of this formulation is that all of
the SU$\left( 2,2|4\right)$ representations that describe Kaluza-Klein
towers in AdS$_{5}\times $S$^{5}$ or $M^{4}\times R^{6}$ supergravity
universally have vanishing eigenvalues for \textit{all} the Casimir
operators. This prediction has been partially verified directly in AdS$%
_{5}\times $S$^{5}$ supergravity. This suggests that the supergravity
spectrum supports a hidden $\left( 10,2\right) $ structure. A possible
duality between AdS$_{5}\times $S$^{5}$ and $M^{4}\times R^{6}$
supergravities is also indicated. Generalizations of the approach applicable
10-dimensional super Yang Mills theory and 11-dimensional M-theory are
briefly discussed.
\end{abstract}

\maketitle

\noindent{CITUSC/02-025 \ \hfill\ \ hep-th/0208012}\bigskip

\bigskip

\address{\large{Department of Physics and Astronomy\\
 University of Southern California,
Los Angeles, CA 90089-0484}} %\date{}

\section{Introduction}

A standard way of describing the motion of a particle on a sphere S$^{5}$ is
to consider the six Euclidean coordinates $X^{I}\left( \tau \right) $ with
the constraint $X^{I}\left( \tau \right) X_{I}\left( \tau \right)
=R_{1}^{2}, $ where the radius $R_{1}$ is a constant in $\tau .$ Similarly
the motion of a particle on AdS$_{5}$ space is described by six coordinates $%
X^{m}\left( \tau \right) $ with $\left( 4,2\right) $ signature with the
constraint $X^{m}X_{m}=-R_{2}^{2}.$ When the constant radii are the same $%
R_{1}=R_{2}=R,$ the combined 12-dimensional space $X^{M}=\left(
X^{m},X^{I}\right) $ with two constraints describes the motion of the
particle on AdS$_{5}\times $S$^{5}.$ One of the constraints on the
12-dimensional dynamical space $X^{M}\left( \tau \right) $ with $\left(
10,2\right) $ signature can be rewritten as a SO$\left( 10,2\right) $
invariant on the light-cone%
\begin{equation}
X^{M}X_{M}=X^{m}X_{m}+X^{I}X_{I}=0.
\end{equation}%
The second constraint (which requires the AdS$_{5}\times $S$^{5}$ radius $R$
to be a constant in $\tau $) will soon be argued to arise from fixing a
local $\left( \tau \text{-dependent}\right) $ gauge symmetry that acts on $%
X^{M}\left( \tau \right) .$

In AdS$_{5}\times $S$^{5}$ supergravity the spectrum is described by
demanding that the eigenvalues of the AdS$_{5}$ Laplacian are correlated
with those of the S$^{5}$ Laplacian, $\nabla _{AdS_{5}}^{2}\psi =-\nabla
_{S_{5}}^{2}\psi ,$ such that the total AdS$_{5}\times $S$^{5}$ Laplacian
vanishes on the supergravity states $\nabla _{AdS_{5}\times S_{5}}^{2}\psi
=0.$ In a previous paper and in an appendix in this paper it is shown that
these properties follow directly from quantizing the 12-dimensional particle 
$X^{M}\left( \tau \right) $ with the constraints described above, such that
the 12 momenta $P_{M}\left( \tau \right) =\left( P_{m}\left( \tau \right)
,P_{I}\left( \tau \right) \right) $ satisfy an additional SO$\left(
10,2\right) $ invariant constraint on the light-cone 
\begin{equation}
P_{M}P^{M}=P_{m}P^{m}+P_{I}P^{I}=0.
\end{equation}%
Under quantization, the $P_{M}P^{M}=0$ condition reduces to the Laplace
equation $\nabla _{AdS_{5}\times S_{5}}^{2}\psi =0.$ The Poisson brackets of
the two constraints $\left[ X^{M}X_{M},P_{M}P^{M}\right] $ shows that there
must be a third SO$\left( 10,2\right) $ invariant constraint%
\begin{equation}
X^{M}P_{M}=X^{m}P_{m}+X^{I}P_{I}=0,
\end{equation}%
This becomes a differential operator in the quantum version which fixes the
overall dimension of the field in 12D.

These three first class constraints $X^{2},P^{2},\left( X\cdot P+P\cdot
X\right) $ close under commutation at the quantum level to form the algebra
of Sp$\left( 2,R\right) $ which is the fundamental gauge symmetry in
two-time physics (2T-physics) \cite{old2T}-\cite{survey2T}. The reader is
also reminded of Dirac's approach \cite{dirac} to conformal symmetry which
consists of differential equations that are equivalent to imposing these
three constraints (indeed AdS$_{5}\times $S$^{5}$ is conformal invariant 
\cite{maldacena}). It is the Sp$\left( 2\right) $ gauge symmetry that allows
the constant radius $R$ of AdS$_{5}\times $S$^{5}$ to be taken independent
of $\tau $ as a gauge choice, thus creating the AdS$_{5}\times $S$^{5}$
phase space out of the flat 12D phase space (this was shown in detail in the
third reference in \cite{old2T} and will be discussed again below). The
gauge can also be fixed to create other 1-time 10-dimensional phase spaces
related to the 12-dimensional phase space, but in this paper we will
concentrate mainly on the AdS$_{5}\times $S$^{5}$ and $M^{4}\times R^{6}$
gauges\footnote{%
The two times are not arbitrarily introduced in 2T-physics. They are a
consequence of the fundamental Sp$\left( 2,R\right) $ phase space local
symmetry. The Sp$\left( 2,R\right) $ is a generalization of $\tau $%
-reparametrization. For a free particle, with only $\tau $-reparametrization
of the worldline, one obtains the constraint $p^{2}=0.$ This constraint is
nontrivial and physical only when there is one timelike dimension. To see
this, consider the worldline formalism for any signature. For Euclidean
signature the solution of $p^{2}=0$ is trivial. If the signature has more
than one time the theory is non-unitary and non-causal since $\tau $%
-reparametrization alone is insufficient to remove ghosts for more than one
time. Hence in a theory with only $\tau $-reparametrization the target
spacetime must have one time, no more and no less. Similarly, Sp$\left(
2,R\right) $ local symmetry, in flat space, requires the three constraints $%
X^{2}=P^{2}=X\cdot P=0,$ which indicate that the physical phase space is
realized only by Sp$\left( 2,R\right) $ gauge invariants. Euclidean or
one-time spacetimes have trivial content, while more than two times lead to
ghosts. Therefore the only non-trivial and physical solution of the Sp$%
\left( 2,R\right) $ constraints requires two timelike dimensions. These
constraints have generalizations in the presence of interactions with
arbitrary background fields, including curved spacetimes \cite{emgrav}\cite%
{highspin}. 2T-physics has the virtue of making hidden symmetries manifest
while unifying various dynamics in 1T-physics as explained in footnote \ref%
{othergauges}.\label{2T}}.

The argument above begins to show that AdS$_{5}\times $S$^{5}$ or other
forms of supergravity may have a hidden 12-dimensional structure. In this
paper we extend these considerations to 12-dimensional super phase space $%
\left( X^{M},P_{M},\Theta \right) $ with 32 fermionic coordinates $\Theta $.
This will be done by studying a 12-dimensional superparticle in the
2T-physics formalism. The results will reveal that AdS$_{5}\times $S$^{5}$
or $M^{4}\times R^{6}$ supergravity indeed have hitherto unnoticed
supersymmetric properties that are related to the 12-dimensional structure
and associated hidden symmetries.

The standard massless superparticle actions in $d=3,4,5,6$ dimensions with $%
N $ supersymmetries have hidden superconformal symmetries, OSp$\left(
N|4\right) $, SU$\left( 2,2|N\right) $, F$\left( 4\right) $, OSp$\left(
8^{\ast }|N\right) $ respectively, whose non-linear realizations are rather
intricate. These hidden symmetries were initially discovered for $d=3,4$ a
long time ago in \cite{schwarz} while the additional cases $d=5,6$ were
first discovered using the two-time physics formulation (2T-physics) \cite%
{old2T}-\cite{survey2T} of the superparticle \cite{super2t}. The non-linear
superconformal symmetries for all cases above are understood much more
simply as linearly realized supersymmetries in $d+2$ dimensions in
2T-physics. The non-linear version emerges after choosing Sp$\left(
2,R\right) $ gauges and expressing the 2T theory in terms of fewer (gauge
fixed) degrees of freedom suitable for one-time physics (1T-physics).

In the 2T approach the theory is formulated in terms of phase space degrees
of freedom in $d+2$ dimensions $\left( X^{M}\left( \tau \right) ,P^{M}\left(
\tau \right) \right) $ and a supergroup element $g\left( \tau \right)
\subset G.$ The supergroup $G$ contains the bosonic subgroup SO$\left(
d,2\right) $ in the spinor representation. A worldline action that is
invariant under linearly realized global supersymmetry $G,$ local Sp$\left(
2,R\right) ,$ local SO$\left( d,2\right) $, and local kappa supersymmetries,
was constructed \cite{super2t}. It was shown that for the special cases of $%
\left( d,G\right) $ dimensions and supergroups $\left( d=3,\,\text{OSp}%
\left( N|4\right) \right) $, $\left( d=4,\,\text{SU}\left( 2,2|N\right)
\right) $, $\left( d=5,\,\text{F}\left( 4\right) \right) $, $\left( d=6,\,%
\text{OSp}\left( 8^{\ast }|N\right) \right) ,$ the 2T action is re-expressed
as the standard superparticle action in 1T-physics after choosing some
special gauges.

In this paper the SU$\left( 2,2|4\right) $ 2T superparticle action will be
extended to a $\left( 10,2\right) $ superparticle action by adding 6 more
coordinates. The 12 coordinates $\left( X^{M}\left( \tau \right)
,P^{M}\left( \tau \right) \right) $ have a special coupling to $g\left( \tau
\right) \subset $SU$\left( 2,2|4\right) $ that respect the global symmetry SU%
$\left( 2,2|4\right) .$ In a special gauge the system describes a
superparticle moving in the 10-dimensional AdS$_{5}\times $S$^{5}$
superspace. The quantum states of this superparticle correspond to the
complete set of the Kaluza-Klein towers that emerge in the AdS$_{5}\times $S$%
^{5}$ compactification of 10D type IIB supergravity.

In section 2, we present the action and discuss its symmetries. In section 3
first class constraints that follow from the gauge symmetries are used to
make gauge invariant statements about the SU$\left( 2,2|4\right) $
representation content of the physical states at the quantum level. It is
shown that the eigenvalues of \textit{all} Casimir operators must vanish in
the physical sector. In section 4 the gauge choice that corresponds to AdS$%
_{5}\times $S$^{5}$ or M$^4\times$R$^6$ is discussed and the quantum states
are given as a superfield that corresponds to the Kaluza-Klein spectrum of
compactified type IIB supergravity. We now outline the essential ideas in
these sections.

The global symmetry of the total action is SU$(2,2|4)$ and the local
symmetries are SO$\left( 4,2\right) \times $SO$\left( 6\right) $, fermionic
kappa, and Sp$\left( 2,R\right) $. These arise as follows. We start with the
standard 2T-physics action in \textit{flat} spacetime that contains the Sp$%
\left( 2,R\right) $ doublets $\left( X^{M}\left( \tau \right) ,P^{M}\left(
\tau \right) \right) $ and the Sp$\left( 2,R\right) $ gauge fields $%
A^{ij}\left( \tau \right) $ that insure a local Sp$\left( 2,R\right) $
symmetry. As usual in 2T-physics, this structure exists non-trivially and is
physically meaningful only when the 12-dimensional spacetime has (10,2)
signature. Hence, this purely bosonic part of the action (first term in Eq.(%
\ref{LL}) below) necessarily has a global SO$\left( 10,2\right) $ symmetry
that acts on the spacetime indices $M.$

Next, to introduce fermions, we consider the supergroup element $g\left(
\tau \right) \subset $SU$\left( 2,2|4\right) $ that contains fermions $%
\Theta .$ All the bosons in $g\left( \tau \right) $ can be removed by the
gauge symmetries discussed below, but in the presence of the bosons in $%
g(\tau )$ the formulation of the theory is more transparent and more
elegant. On the left and right sides of $g\left( \tau \right) $ we define
the supergroup transformations SU$\left( 2,2|4\right) _{L}\times $SU$%
(2,2|4)_{R}.$ On the left side of $g\left( \tau \right) ,$ and on $\left(
X^{M}\left( \tau \right) ,P^{M}\left( \tau \right) \right) ,$ we gauge the
bosonic subgroup SO$\left( 4,2\right) \times $SO$\left( 6\right) =$SU$\left(
2,2\right) \times $SU$\left( 4\right) $ that sits at the intersection of SO$%
\left( 10,2\right) $ and SU$\left( 2,2|4\right) _{L}.$ This couples $\left(
X^{M},P^{M}\right) $ to $g$ (second term in Eq.(\ref{LL}) below) and thus
breaks the 12D symmetry of the first term down to the SO$\left( 4,2\right)
\times $SO$\left( 6\right) $ gauge symmetry of the total action. The action
also has a subtle fermionic local kappa supersymmetry embedded in SU$\left(
2,2|4\right) _{L}$ acting on the left side of $g\left( \tau \right) $ and on
the Sp$\left( 2,R\right) $ gauge fields $A^{ij}\left( \tau \right) $.

In addition to the local symmetries, there is a global SU$(2,2|4)_{R}$
supersymmetry of the system that acts on the right side of $g\left( \tau
\right) .$ Initially the global SU$(2,2|4)_{R}$ does not transform $\left(
X^{M},P^{M}\right) $. However, after a gauge fixing that removes the bosons
in $g\left( \tau \right) $ the SU$(2,2|4)_{R}$ induces a transformation on $%
\left( X^{M},P^{M}\right) $ as follows. The gauged fixed $g$ takes the form 
\begin{equation}
g=\exp \left( 
\begin{array}{cc}
0 & \Theta \\ 
\bar{\Theta} & 0%
\end{array}%
\right) .  \label{gfixed}
\end{equation}%
The $\Theta $ are in $\left( 4,\bar{4}\right) $ spinor representation of SO$%
\left( 4,2\right) \times $SO$\left( 6\right) .$ A global SU$(2,2|4)_{R}$
transformation of the gauged fixed $g$ must be compensated from the left by $%
\Theta \left( \tau \right) $-dependent local SO$\left( 4,2\right) \times $SO$%
\left( 6\right) $ transformations in order to maintain the gauge fixed form.
However, the $\Theta $-dependent local SO$\left( 4,2\right) \times $SO$%
\left( 6\right) $ rotates also $\left( X^{M},P^{M}\right) .$ This
combination of global and local transformations become the expected
spacetime supersymmetry in 12D super phase space $\left( X,P,\Theta \right) $%
, but it is important to understand that its origin is the much simpler SU$%
(2,2|4)_{R}$ acting only on $g\left( \tau \right) .$

Thus the local symmetries are Sp$\left( 2,R\right) ,$ SO$\left( 4,2\right)
\times $SO$\left( 6\right) $, and the fermionic kappa, while the global
symmetry is SU$(2,2|4)_{R}$, all of which are linearly realized in the 2T
theory described by the deceptively simple invariant action in Eq.(\ref{LL})
below.

In section 3 an interesting set of constraints is obtained on the Noether
charges $J$ of the global SU$(2,2|4)_{R}$ given in Eq.(\ref{J}). These
constraints imply that only certain special representations of SU$\left(
2,2|4\right) $ can be realized on physical states. In particular \textit{all}
the SU$\left( 2,2|4\right) $ Casimir eigenvalues Str$\left( J^{n}\right) $
must vanish for all the physical states that satisfy the Sp$\left(
2,R\right) $ 12-dimensional constraints $X^{2}=X\cdot P=P^{2}=0.$ The same
zero Casimirs must occur in AdS$_{5}\times $S$^{5}$ supergravity since the
physical states coincide with the Kaluza-Klein towers that emerge in the AdS$%
_{5}\times $S$^{5}$ compactification spectrum of type-IIB supergravity. The
same conclusion applies to other forms of supergravity spectra in super
phase spaces (such as on $M^{4}\times R^{6}$) obtainable via gauge fixing of
the 12-dimensional action.

This striking 2T-physics prediction about AdS$_{5}\times $S$^{5}$
supergravity has actually been verified purely within the context of
supergravity. Although it has been known for many years that the standard
oscillator approach for unitary representations of SU$(2,2|4)$ \cite%
{barsgunaydin} describes the Kaluza-Klein towers of supergravity \cite{MG}%
\cite{KRN}, the quadratic Casimir of these infinite set of representations
was not correctly computed. After the universal zero Casimir eigenvalues was
predicted by 2T-physics (in this paper), the direct computation of the
quadratic Casimir Str$\left( J^{2}\right) $ within supergravity \cite{zero}
verified the expected zero result.

The universal zero eigenvalues found here and \cite{zero} are not automatic
consequences of representations of SU$\left( 2,2|4\right) $. Indeed, as
shown in \cite{zero}, among the infinite number of lowest level (singular)
representations of SU$\left( 2,2|4\right) $, most representations have
non-zero quadratic Casimir eigenvalues. Our universal zero is a nontrivial
property of the representations selected by the Sp$\left( 2,R\right) $ gauge
invariance conditions $X^{2}=X\cdot P=P^{2}=0$, and is explained
fundamentally by the 12-dimensional structure of the theory presented here.

It is important to emphasize that the physical constraints on the SU$\left(
2,2|4\right) $ representations are fully covariant under all symmetries of
the model, and are independent of any gauge fixing. AdS$_{5}\times $S$^{5}$
supergravity happens to emerge in one of the gauge choices, and therefore it
shares the same SU$\left( 2,2|4\right) $ properties with any other
1T-physics model that would emerge from a different choice of gauge.

In section 4 we show how the AdS$_{5}\times $S$^{5}$ superparticle action
emerges by a series of gauge choices. First using the SO$\left( 4,2\right)
\times $SO$\left( 6\right) $ gauge symmetry we remove all the bosons from $%
g\left( \tau \right) $ and keep only the fermions $\Theta \left( \tau
\right) $ as in Eq.(\ref{gfixed}). In this gauge the remaining degrees of
freedom are described by the 12-dimensional phase space superspace $\left(
X^{M},P^{M},\Theta \right) $ with 32 real (or 16 complex) fermionic
components in $\Theta $. Before choosing such a gauge the 2T action had the
global supersymmetry SU$(2,2|4)_{R}$ as a linearly realized supersymmetry
acting on the right side of $g\left( \tau \right) $, but it did not act on $%
\left( X^{M},P^{M}\right) .$ After choosing the gauge in Eq.(\ref{gfixed})
the global SU$\left( 2,2|4\right) $ supersymmetry becomes non-linearly
realized and correctly gives the supertransformations of the 12D superspace $%
\left( X^{M},P^{M},\Theta \right) $ as already mentioned above.

Next, one can make gauge choices for Sp$\left( 2,R\right) $ and the
fermionic local kappa supersymmetries that are still local symmetries of the
action constructed from the 12-D superspace $\left( X^{M},P^{M},\Theta
\right) $. These gauge symmetries reduce the 2T degrees of freedom from the
12D superspace to 10-dimensional AdS$_{5}\times $ S$^{5}$ or $M^{4}\times
R^{6}$ superspace with one time. In principle, there are many possible gauge
choices that produce various 10D superspaces, but only two of these, the AdS$%
_{5}\times $S$^{5}$ or $M^{4}\times R^{6}$ gauges will be explored in this
paper. These choices produce the superspaces for a 10D super particle moving
on AdS$_{5}\times $S$^{5}$ or $M^{4}\times R^{6}.$ After quantization, the
physical states of these 1T systems coincide with the spectrum of the
corresponding form of supergravity. The spectrum is simply described by a
single superfield that provides a basis for a nonlinearly realized unitary
representation of the global supersymmetry SU$(2,2|4)_{R}$. The non-linearly
realized generators of SU$(2,2|4)_{R}$ that act on the superfield are
explicitly given in terms of the canonical degrees of freedom in a 1T
formalism. They are fairly complicated expressions, but they are derived
directly from the 12D action as the Noether charges $J$ that have a very
simple and compact expression in the 2T formalism.

Other gauge fixed forms of the same 2T action can be considered\footnote{%
The 2T formulation of the theory is more than a trivial embedding in higher
dimensions that obey constraints. The non-triviality comes from the fact
that the 2T theory describes not only one, but many one-time dynamical
systems. The one-time systems form a family of dual theories that
holographically represent the same $d+2$ theory in many ways in $d$
dimensions. The different 1T dynamics come about through Sp$\left(
2,R\right) $ gauge fixing that identifies `time\textquotedblright . Choices
of gauges in 2T-physics usually identify the proper time $\tau $ with some
combination of the two timelike coordinates $X^{0}\left( \tau \right)
,X^{0^{\prime }}\left( \tau \right) $ in target spacetime. Each such gauge
choice corresponds to a different rearrangement of the time evolution of the
remaining target space degrees of freedom as a function of the chosen
\textquotedblleft time\textquotedblright\ in target space. Therefore a
family of different Hamiltonians (hence different 1T-dynamics) are obtained
by gauge fixing a given 2T-physics theory. There are measurable consequences
of the relationships among the 1T systems that emerge from the same
2T-theory. The simplest example of such relations is the same Casimirs and
representations of the global symmetry, as already mentioned in this paper,
but there is much more content that can be tested. The success of such tests
for a few simple quantum mechanical systems \cite{old2T} have already
established that the 2T approach has measurable physical consequences. This
higher dimensional unification of apparently different 1T- physics dynamics
(or \textquotedblleft dually\textquotedblright\ related 1T systems), that
really describe the same 2T-system, is one of the advantages of 2T-physics. %
\label{othergauges}}. As is usual in 2T physics, there are many gauges that
give different 1T dynamical systems which form a family of dual systems.
This provides new realizations of SU$\left( 2,2|4\right) $ in 1T systems
that are expected to be \textquotedblleft dual\textquotedblright\ to the AdS$%
_{5}\times $S$^{5}$ supergravity system. We will make comments about more
general gauge choices at various points in the paper, but the only gauges
that will be considered in detail in the present paper are the AdS$%
_{5}\times $S$^{5}$ or $M^{4}\times R^{6}$ gauges.

\section{Action and symmetries}

We define 12 coordinates $X^{M}$ in flat spacetime with $\left( 10,2\right) $
signature and divide them into two sets $X^{M}=\left( X^{m},X^{I}\right) .$
The $\left( 10,2\right) $ signature is not chosen \textquotedblleft by
hand\textquotedblright , rather it is required as a consequence of the local
Sp$\left( 2,R\right) $ symmetry$^{\ref{2T}}$. Similarly we define the 12
momenta $P_{M}=\left( P_{m},P_{I}\right) .$ We indicate $X_{1}^{M}\equiv
X^{M}$ and $X_{2}^{M}=P^{M},$ and use the notation $X_{i}^{M}$ with $i=1,2$
to indicate that the 12 coordinates and momenta form doublets of Sp$\left(
2,R\right) $ for every spacetime index $M.$ The orbital \textquotedblleft
angular momenta\textquotedblright\ in the 12-dimensional space are Sp$\left(
2,R\right) $ gauge invariant singlets 
\begin{equation}
L^{MN}=X^{M}P^{N}-X^{N}P^{M}=\varepsilon ^{ij}X_{i}^{M}X_{j}^{N}.
\end{equation}%
The $L^{MN}$ are the generators of SO$\left( 10,2\right) $ that acts on the
spacetime indices $M.$ The subset $L^{mn},L^{IJ}$ \ are the generators of
the subgroup SO$\left( 4,2\right) \times $SO$\left( 6\right) $ which will
later be related to the Killing symmetry of the space AdS$_{5}\times $S$%
^{5}. $

We also introduce the supergroup element $g\left( \tau \right) \in $SU$%
\left( 2,2|4\right) .$ This contains bosonic and fermionic degrees of
freedom. The bosons are in the adjoint representation of SO$\left(
4,2\right) \times $SO$\left( 6\right) =$SU$\left( 2,2\right) \times $SU$%
\left( 4\right) ;$ these will later be eaten away by a SO$\left( 4,2\right)
\times $SO$\left( 6\right) $ gauge symmetry that sits at the intersection SO$%
\left( 10,2\right) \cap $SU$\left( 2,2|4\right) $. The fermions $\Theta
\left( \tau \right) $ are in the $\left( 4,\bar{4}\right) $ representation
of SU $\left( 2,2\right) \times $SU$\left( 4\right) $ (i.e. spinor $\otimes $
spinor of SO$\left( 4,2\right) \times $SO$\left( 6\right) $).

The 2T Lagrangian with an invariant coupling among these degrees of freedom
is 
\begin{equation}
\mathcal{L}=\left( \dot{X}_{1}^{M}X_{2}^{N}-\frac{1}{2}%
A^{ij}X_{i}^{M}X_{j}^{N}\right) \eta _{MN}-\frac{1}{16}Str\left( L\left(
\partial _{\tau }gg^{-1}\right) \right) ,  \label{LL}
\end{equation}%
where 
\begin{equation}
L=\left( 
\begin{array}{cc}
L^{mn}\Gamma _{mn} & 0 \\ 
0 & -L^{IJ}\Gamma _{IJ}%
\end{array}%
\right)  \label{Lmatrix}
\end{equation}%
$\eta _{MN}$ is the flat metric with $\left( 10,2\right) $ signature. The
gamma matrices $\left( \Gamma _{mn},\Gamma _{IJ}\right) $ are given in
detail in Appendix-A. They form the spinor representation of the generators
of SO$\left( 4,2\right) \times $SO$\left( 6\right) =$ SU$\left( 2,2\right)
\times $SU$\left( 4\right) $. The negative sign in the lower block in (\ref%
{Lmatrix}) is multiplied by another negative sign when the supertrace in Eq.(%
\ref{LL}) is evaluated. Thus the coupling consists of projecting the Cartan
connection $\partial _{\tau }gg^{-1}$ to the subgroup SU$\left( 2,2\right)
\times $SU$\left( 4\right) $ and then coupling it to the SO$\left(
4,2\right) \times $SO$\left( 6\right) $ orbital \textquotedblleft angular
momenta\textquotedblright\ $\left( L^{mn},L^{IJ}\right) .$ Note that the
projected Cartan connection is \textit{not a pure gauge} since it includes
contributions from the coset parameters $\Theta .$

This action is a generalization of the superparticle action in $d=4$ with SO$%
\left( 4,2\right) $ superconformal symmetry in the 2T formalism \cite%
{super2t}. In that case we had only the subset of 6 coordinates and momenta $%
\left( X^{m},P^{m}\right) $ with a coupling to $g\left( \tau \right) $ that
corresponds to the upper block in the matrix in (\ref{Lmatrix}). In the
present case we have the additional 6 coordinates and momenta $X^{I},P^{I}$
with a coupling to $g\left( \tau \right) $ via the lower block in the matrix.

In the absence of $g\left( \tau \right) $ the purely bosonic Lagrangian is
invariant under a global SO$\left( 10,2\right) .$ This SO$\left( 10,2\right) 
$ is unavoidable because the correct normalization of the kinetic term $\dot{%
X}\cdot P$ imposes it automatically and the \textit{non-Abelian} symmetry Sp$%
\left( 2,R\right) $ forces the same $\eta _{MN}$ on the quadratic dot
products $X_{i}\cdot X_{j}$ that are the coefficients of the gauge field $%
A^{ij}.$ Thus the dot product cannot be split into two parts with different
coefficients for the $\left( 4,2\right) $ and $\left( 6,0\right) $ parts.
Therefore the 12-dimensional constraints 
\begin{equation}
X^{2}=P^{2}=X\cdot P=0  \label{constr}
\end{equation}%
that follow from the $A^{ij}$ equations of motion are necessarily invariant
under SO$\left( 10,2\right) .$

Although the first term in the Lagrangian has manifest SO$\left( 10,2\right) 
$ symmetry, the second term breaks it to manifest subgroup symmetry SO$%
\left( 4,2\right) \times $SO$\left( 6\right) ,$ which is also the bosonic
subgroup in SU$\left( 2,2|4\right) _{L}.$ We will see below that the common
subgroup that sits at the intersection of SU$\left( 2,2|4\right) \cap $SO$%
\left( 10,2\right) $ is a \textit{local }SO$\left( 4,2\right) \times $SO$%
\left( 6\right) $ \textit{space-time Lorentz symmetry}. The relative
coefficient $\left( -1/16\right) $ between the two terms in $\mathcal{L}$\
is fixed by this local symmetry. We see that there are no free parameters in
this action.

The action has a number of global and local symmetries. The brief
description below summarizes the essential aspects of more detailed symmetry
discussions given in \cite{super2t}.

(1) From the extensive analysis in \cite{old2T}-\cite{survey2T} we already
know that the Lagrangian above has \textit{local} symmetry under Sp$\left(
2,R\right) $ separately for both terms in the Lagrangian ($X_{i}^{M}$ is a
doublet, $A^{ij}$ is a triplet gauge field, $g$ is a singlet, $L^{mn}$ and $%
L^{IJ}$ are invariant). This is the basic local symmetry of 2T-physics; it
is responsible for the constraints $X_{i}\cdot X_{j}=0$ that follow as the
equations of motion for the gauge field $A^{ij}.$ The solution of these
constraints requires two time-like coordinates. This local symmetry is also
responsible for removing the ghosts associated with two timelike dimensions.
No more and no less than two timelike coordinates are possible for a
non-trivial unitary system consistent with local Sp$\left( 2,R\right) $.
This symmetry allows one to recast the 2T system in fixed gauges in the form
of many different looking 1T dynamical systems (different Hamiltonians \cite%
{old2T}) that are related to each other by duality type Sp$\left( 2,R\right) 
$ transformations. One of these gauge choices will lead to the AdS$%
_{5}\times $ S$^{5}$ Kaluza-Klein towers.

(2) The Cartan connection $\partial _{\tau }gg^{-1}$ is invariant under the
transformation of $g\left( \tau \right) $ by \textit{right multiplication} $%
g\left( \tau \right) \rightarrow g\left( \tau \right) g_{R},$ for \textit{%
global} $g_{R}\subset $SU$\left( 2,2|4\right) _{R}$ . Therefore, with $%
X_{i}^{M}$ and $A^{ij}$ taken as singlets under SU$\left( 2,2|4\right) _{R}$
the Lagrangian is invariant under 
\begin{equation}
g\rightarrow gg_{R}.  \label{global}
\end{equation}%
Via Noether's theorem, the Lagrangian yields the conserved charges $J\left(
\tau \right) $ which can be written in the form of a 8$\times 8$ supermatrix
that belongs to the Lie algebra of SU$\left( 2,2|4\right) _{R}$ 
\begin{equation}
J=\frac{i}{4}g^{-1}Lg,\quad \partial _{\tau }J=0.  \label{J}
\end{equation}%
The conservation of the charges $\partial _{\tau }J=0$ is verified by using
the equations of motion. This global supersymmetry is physical since it is
gauge invariant, and cannot be altered by making gauge choices. Therefore,
no matter what gauge is chosen to express the theory in terms of canonical
variables, the SU$\left( 2,2|4\right) _{R}$ bosonic and fermionic charges
described by $J$ must remain conserved, and must form the Lie superalgebra
of SU$\left( 2,2|4\right) _{R}$ under Poisson brackets at the classical
level, or under supercommutators when the theory is quantized and operators
properly ordered. This observation will play a crucial role in understanding
the supersymmetry SU$\left( 2,2|4\right) $ in terms of AdS$_{5}\times $S$%
^{5} $ superspace$,$ supertwistors, or other degrees of freedom arrived at
by a variety of gauge fixing choices.

(3) There is a bosonic \textit{local} Lorentz symmetry SO$\left( 4,2\right)
\times $SO$\left( 6\right) $ defined by the intersection of SO$\left(
10,2\right) \cap $SU$\left( 2,2|4\right) _{L}.$ The local group element $%
g_{L}\left( \tau \right) \in $ SU$\left( 2,2|4\right) _{L}$ acts on $g\left(
\tau \right) $ by \textit{left multiplication }$g\rightarrow g_{L}g$\textit{%
. }The local symmetry has infinitesimal parameters $\omega ^{mn}\left( \tau
\right) ,\omega ^{IJ}\left( \tau \right) .$ which act on $g$ in the spinor
representation and on $X_{i}^{m},X_{i}^{I}$ in the vector representation 
\begin{eqnarray}
\delta _{\omega }X_{i}^{m} &=&\omega ^{mn}\left( X_{i}\right) _{n},\quad
\delta _{\omega }X_{i}^{I}=\omega ^{IJ}\left( X_{i}\right) _{J},\quad \\
\delta _{\omega }g &=&\frac{1}{4}\left( 
\begin{array}{cc}
\omega ^{mn}\Gamma _{mn} & 0 \\ 
0 & \omega ^{IJ}\Gamma _{IJ}%
\end{array}%
\right) \,g,\quad \delta _{\omega }A^{ij}=0.
\end{eqnarray}%
The Lagrangian $\mathcal{L}$ and the physical SU$\left( 2,2|4\right) _{R}$
current $J$ are both invariant. The time derivatives $\partial _{\tau
}\omega ^{mn}\left( \tau \right) ,$ $\partial _{\tau }\omega ^{IJ}\left(
\tau \right) $ produced by the two terms in the Lagrangian cancel each
other. This is what fixes the relative coefficient $-1/16$ in the
Lagrangian. It is possible to display this symmetry more clearly by
rewriting the Lagrangian in the form (up to a total $\tau $ derivative) 
\begin{equation}
\mathcal{L}=\frac{1}{2}\left( D_{\tau }X_{i}\right) \cdot X_{j}\varepsilon
^{ij},\quad  \label{LLL}
\end{equation}%
where $D_{\tau }X_{i}^{M}$ is the covariant derivative under both Sp$\left(
2,R\right) $ and SO$\left( 4,2\right) \times $SO$\left( 6\right) $ local
transformations 
\begin{equation}
D_{\tau }X_{i}^{M}=\partial _{\tau }X_{i}^{M}-\varepsilon
_{ik}A^{kj}X_{j}^{M}-\Omega _{\,\,N}^{M}X_{i}^{N}
\end{equation}%
constructed by using the Cartan connection $\Omega ^{MN}$ for the bosonic
subgroup in SU$\left( 2,2|4\right) _{L}$%
\begin{eqnarray}
\Omega ^{MN} &=&\frac{1}{16}Str\left( \Gamma ^{MN}\partial _{\tau
}gg^{-1}\right) ,\quad \\
\Gamma ^{MN} &=&\left( 
\begin{array}{cc}
\Gamma ^{mn} & 0 \\ 
0 & 0%
\end{array}%
\right) ,\left( 
\begin{array}{cc}
0 & 0 \\ 
0 & -\Gamma ^{IJ}%
\end{array}%
\right) .
\end{eqnarray}%
Note that $\Omega ^{mI}$ or $\Gamma ^{mI}$ are absent.

(4) There is also a local fermionic kappa supersymmetry embedded in SU$%
\left( 2,2|4\right) _{L}.$ The fermionic coset elements in SU$\left(
2,2|4\right) _{L}$ act on $g$ from the left for infinitesimal $K$ as follows 
\begin{eqnarray}
\delta _{\kappa }g &=&Kg,\quad \delta _{\kappa }A^{ij}\neq 0,\quad \delta
_{\kappa }X_{i}^{M}=0,\quad \\
K &=&\left( 
\begin{array}{cc}
0 & \xi \left( \tau \right) \\ 
\tilde{\xi}\left( \tau \right) & 0%
\end{array}%
\right) .
\end{eqnarray}%
Here $\delta _{\kappa }A^{ij}$ is non-zero as specified below. The local
fermionic parameter $\xi \left( \tau \right) ,$ which is classified as $%
\left( 4,\bar{4}\right) $ under the subgroup SU$\left( 2,2\right) \times $SU$%
\left( 4\right) $ must take the form (for reasons explained below) 
\begin{equation}
\xi \left( \tau \right) =\varepsilon ^{ij}X_{i}^{m}X_{j}^{I}\left( \Gamma
_{m}\kappa \Gamma _{I}\right) =L^{mI}\,\left( \Gamma _{m}\kappa \Gamma
_{I}\right) \,,  \label{xi}
\end{equation}%
where the local $\kappa \left( \tau \right) $ are the unrestricted local
fermionic parameters also classified as $\left( 4,\bar{4}\right) $. Under
such a transformation we obtain after some simplification 
\begin{equation}
\delta _{\kappa }\mathcal{L}=-\frac{1}{2}\left( \delta _{\kappa
}A^{ij}\right) X_{i}\cdot X_{j}+\frac{1}{8}Str\left( \left( 
\begin{array}{cc}
0 & \psi \left( \xi \right) \\ 
\tilde{\psi}\left( \xi \right) & 0%
\end{array}%
\right) \left( \partial _{\tau }gg^{-1}\right) \right) ,  \label{delkappa}
\end{equation}%
where $\psi \left( \xi \right) \equiv L^{mn}\left( \Gamma _{mn}\xi \right)
+L^{IJ}\left( \xi \Gamma _{IJ}\right) .$ The only way to achieve kappa
invariance $\delta _{\kappa }\mathcal{L}=0$ is to arrive at a $\psi \left(
\xi \right) $ that is proportional to $X_{i}\cdot X_{j}$ so that $\delta
_{\kappa }A^{ij}$ can be chosen to cancel the contribution from the second
term. Indeed, this property is satisfied for arbitrary $\kappa \left( \tau
\right) $ provided $\xi \left( \tau \right) $ is of the form given in (\ref%
{xi}). To prove this we insert the form of $\xi \left( \tau \right) $ in $%
\psi \left( \xi \right) ,$ work out the algebra of the gamma matrices by
using $\Gamma _{MN}\Gamma _{R}=\Gamma _{MNR}+\eta _{NR}\Gamma _{M}-\eta
_{MR}\Gamma _{N}$ (for $M=m,I$ etc.) and note that the antisymmetric $\Gamma
_{MNR}$ term forces antisymmetry in the indices $\left[ mnr\right] $ or $%
\left[ IJK\right] $ for the expressions $X_{i}^{[m}X_{j}^{n}X_{k}^{r]}$ or $%
X_{i}^{[I}X_{j}^{J}X_{k}^{K]}$ that appear in $\psi \left( \xi \right) .$
These terms vanish due to the fact that the Sp$\left( 2,R\right) $ indices $%
i,j,k$ take only two possible values. The remaining terms in $\psi \left(
\xi \right) $ generate dot products $X_{i}^{m}X_{j}^{n}\eta _{mn}$ or $%
X_{i}^{I}X_{j}^{J}\delta _{IJ}$ in such a way that they assemble to the 
\textit{total} 
\begin{equation}
X_{i}\cdot X_{j}\equiv X_{i}^{m}X_{j}^{n}\eta _{mn}+X_{i}^{I}X_{j}^{J}\delta
_{IJ},  \label{correct}
\end{equation}%
with the \textit{correct relative coefficient consistent with }SO$\left(
10,2\right) $. Therefore $\psi \left( \xi \right) $ is proportional to the SO%
$\left( 10,2\right) $ invariant $X_{i}\cdot X_{j}$ when $\xi \left( \tau
\right) $ is of the form (\ref{xi}). This last point is essential to have
kappa supersymmetry since the coefficient of $\left( \delta _{\kappa
}A^{ij}\right) $ in Eq.(\ref{delkappa}) is necessarily invariant under SO$%
\left( 10,2\right) $ as noted in the paragraph following Eq.(\ref{LL}). We
can then choose $\delta _{\kappa }A^{ij}$ to cancel the coefficient of $%
X_{i}\cdot X_{j}$ produced by the second term in $\delta _{\kappa }\mathcal{L%
}$ in Eq.(\ref{delkappa}) and obtain kappa supersymmetry, off-shell. The
physical current $J$ in (\ref{global}) is also invariant under local kappa
transformations in the physical sector $\delta _{\kappa }J=0$. This is shown
by using the same arguments, but now imposing the on shell condition $%
X_{i}\cdot X_{j}\equiv 0,$ which is true for the Sp$\left( 2,R\right) $
gauge invariant sector (physical states).

\section{Gauge invariant constraints}

There are three constraint equations $X_{i}\cdot X_{j}\equiv
X_{i}^{M}X_{j}^{N}\eta _{MN}=0$ that follow from the equations of motion for 
$A^{ij}$ 
\begin{equation}
X_{i}\cdot X_{j}=X_{i}^{m}X_{j}^{n}\eta _{mn}+X_{i}^{I}X_{j}^{J}\delta
_{IJ}=0.  \label{xixj}
\end{equation}%
It will be useful to write them more explicitly in the form 
\begin{equation}
X_{m}X^{m}=-X_{I}X^{I},\,\,P_{m}P^{m}=-P_{I}P^{I},\,\,X_{m}P^{m}=-X_{I}P^{I}
\label{xixj2}
\end{equation}%
The expressions $X_{i}\cdot X_{j}$ are the three generators of Sp$\left(
2,R\right) $ as seen by commuting them. Their vanishing implies that the
physical phase space or the physical states of the theory are Sp$\left(
2,R\right) $ singlets.

Let us examine the gauge invariant SU$\left( 2,2|4\right) _{R}$ charges $J.$
The square of this 8$\times 8$ supermatrix gives (taking into account
quantum ordering of operators) 
\begin{equation}
\left( J\right) ^{2}=-\frac{1}{16}g^{-1}\left( L\right) ^{2}g=\frac{1}{4}%
g^{-1}\left( 
\begin{array}{cc}
C_{2}^{\left( 4,2\right) } & 0 \\ 
0 & C_{2}^{\left( 6,0\right) }%
\end{array}%
\right) g-2\hbar \left( J\right)
\end{equation}%
where $C_{2}^{\left( 4,2\right) }$, $C_{2}^{\left( 6,0\right) }$ are the
Casimir operators of the orbital SO$\left( 4,2\right) \times $SO$\left(
6\right) $ subgroup%
\begin{equation}
C_{2}^{\left( 4,2\right) }=\frac{1}{2}L_{mn}L^{mn},\quad C_{2}^{\left(
6,0\right) }=\frac{1}{2}L_{IJ}L^{IJ}
\end{equation}%
If we examine the constraints (\ref{xixj2}) we find that these two Casimirs
must be the same (respecting order of quantum operators) 
\begin{align}
C_{2}^{\left( 4,2\right) }& =X_{m}\left( P_{n}P^{n}\right) X^{m}-\left(
X_{m}P^{m}\right) \left( P^{n}X_{n}\right)  \label{c42} \\
& =X_{I}\left( P_{J}P^{J}\right) X^{I}-\left( X_{I}P^{I}\right) \left(
P^{J}X_{I}\right) \\
& =C_{2}^{\left( 6,0\right) }.  \label{c6}
\end{align}%
Therefore the first term in $\left( J\right) ^{2}$ is proportional to the
identity supermatrix $\mathbf{1,}$ yielding%
\begin{equation}
\left( J\right) ^{2}=\frac{1}{4}C_{2}^{\left( 6,0\right) }\mathbf{1-}2\hbar
\left( J\right) =\frac{\hbar ^{2}}{4}l\left( l+4\right) \mathbf{-}2\hbar
\left( J\right)  \label{j2}
\end{equation}%
Higher powers $\left( J\right) ^{n}$ are now easily computed by repeatedly
using the formula for $\left( J\right) ^{2}.$ The eigenvalues of $%
C_{2}^{\left( 6,0\right) }=\hbar ^{2}l\left( l+4\right) $ will increase with 
$l=0,1,2,\cdots $, where $l$ labels the harmonics on $S^{5}$ (see
Appendix-B). The representation of SU$\left( 2,2|4\right) _{R}$ changes as $%
l $ changes, therefore this theory describes an infinite number of
representations of the supersymmetry SU$\left( 2,2|4\right) _{R}$
(corresponding to the supergravity Kaluza-Klein towers).

Since the left side of Eq.(\ref{j2}) (and similarly $\left( J\right) ^{n})$
is an 8$\times 8$ supermatrix, this equation expresses rather strong
constraints on the generators of SU$\left( 2,2|4\right) _{R},$ such that
only certain representations of SU$\left( 2,2|4\right) _{R}$ can be realized
in the physical sector of this theory. In particular the quadratic and all
higher Casimir operators of SU$\left( 2,2|4\right) _{R}$ must vanish in
these realizations since the supertrace of $\mathbf{1}$\textbf{\ }and the
supertrace of\textbf{\ }$J$\textbf{\ }are zero 
\begin{equation}
C_{n}^{\left( 2,2|4\right) }=Str\left( J^{n}\right) =0.  \label{Cn}
\end{equation}%
These SU$\left( 2,2|4\right) _{R}$ properties arose through the
12-dimensional constraints $X^{2}=P^{2}=X\cdot P=0,$ hence the
representations of SU$\left( 2,2|4\right) _{R}$ that are selected through
these physical state constraints reflect the underlying 12-dimensional
structure.

Using the local gauge symmetries for kappa, Sp$\left( 2,R\right) $, SO$%
\left( 4,2\right) \times $SO$\left( 6\right) $ described in the previous
section, both the Lagrangian and the SU$\left( 2,2|4\right) _{R}$ charges $J$
can be gauge fixed to various forms and expressed in terms of fewer
(physical) degrees of freedom. The details of a specific gauge fixing will
be given in the next section, but since the model can be reduced to many
possible 1T-physics versions by different gauge choices, it is useful to
first argue in general terms about the effect of any gauge fixing on the
global supersymmetry SU$\left( 2,2|4\right) _{R}$ that commutes with all the
gauge symmetries. Since both $\mathcal{L}$ and $J$ are gauge invariant, the
gauge fixed Lagrangian still has the same SU$\left( 2,2|4\right) _{R}$
global supersymmetry after gauge fixing, but then it is realized nonlinearly
on the remaining fewer degrees of freedom. The conserved charges $J$ for the
nonlinearly realized global supersymmetry SU$\left( 2,2|4\right) _{R}$ are
still given by the same gauge invariant $J$ that appears in Eq.(\ref{J}).
But after gauge fixing, $J$ is expressed in terms of the fewer degrees of
freedom that results from inserting the gauge fixed $X,P,g$ into the
original expression for $J$. This gauge fixed form of $J$ is still conserved 
$\partial _{\tau }J\left( \tau \right) =0$ since this is a gauge invariant
equation. Therefore, in fixed gauges, although the realizations are in terms
of various degrees of freedom, they all represent the same irreducible
representation of SU$\left( 2,2|4\right) _{R}$ that satisfy Eqs.(\ref{j2},%
\ref{Cn}). This symmetry is one of the observables that ties together the
different looking 1T systems.

Typically the SU$\left( 2,2|4\right) $ symmetry is non-linearly realized on
the remaining superspace coming both from the gauge fixed $X_{i}^{M}$ and $g$%
. By contrast, initially the global supersymmetry was linearly realized on
the full $g\rightarrow gg_{R},$ and it did not transform $X^{M}$ before
gauge fixing. The reason that $X^{M}$ must transform after the gauge fixing
is understood as follows: for the gauge fixed $g\left( \tau \right) ,$ the
global transformation from the right $g\rightarrow gg_{R}$ must be followed
by a compensating local transformation SO$\left( 4,2\right) \times $SO$%
\left( 6\right) $ from the left in order to maintain its gauge fixed form,
but the compensating local transformation SO$\left( 4,2\right) \times $SO$%
\left( 6\right) $ must also act on $X_{i}^{M};$ therefore the supersymmetry
generated by the global parameters $g_{R}$ induces transformations on the
entire physical superspace defined after any gauge fixing. The infinitesimal
non-linear transformation rules for SU$\left( 2,2|4\right) $ in the gauge
fixed sectors are automatically obtained by commuting the gauge fixed $J$
with the remaining canonical degrees of freedom. These nonlinear
transformations are guaranteed to be the global SU$\left( 2,2|4\right) _{R}$
symmetries of the gauge fixed action (for similar simpler bosonic examples
see the third reference in \cite{old2T} for detailed computations).

The expressions that we will use below for the AdS$_{5}\times $S$^{5}$ gauge
fixed versions of the charges $J$ or $L^{mn}$ will be needed in this paper
at the classical level, so we will not care about quantum ordering of
non-linear expressions for most of our discussion. However, quantum ordering
of operators must be done such that the vanishing of the quadratic Casimirs $%
C_{n}\left( 2,2|4\right) =0$ must be obeyed as a gauge invariant condition
on the quantum states. We have indeed verified that $C_{2}\left(
2,2|4\right) =0$ is satisfied at the quantum level by the AdS$_{5}\times $S$%
^{5}$ Kaluza-Klein towers in supergravity \cite{zero}.

\section{AdS$_{5}\times$S$^{5}$ Kaluza-Klein towers}

We will choose a gauge that reduces the 2T system above to a 1T system that
describes the AdS$_{5}\times $S$^{5}$ supersymmetric Kaluza-Klein towers. We
will express the model in the fixed gauge in terms of supercoordinates on
the worldline (positions, momenta, theta). Our approach provides a
fundamental description of the system and explains the existence of a
multiplet structure for the entire tower system. This is in agreement with
the Kaluza-Klein towers that result from the AdS$_{5}\times $S$^{5}$
compactification of 10D type IIB supergravity \cite{MG}\cite{KRN}, therefore
we obtain a hidden 12-dimensional interpretation of the supergravity
spectrum of states. We will choose the gauges to reveal the symmetry
structures in stages.

\subsection{12D superspace or twistor gauge (local Lorentz gauge fixing)}

Using the local SO$\left( 4,2\right) \times $SO$\left( 6\right) $ symmetry
contained in the intersection of SO$\left( 10,2\right) \cap G_{L}$ we can
eliminate all the bosons in $g\left( \tau \right) $ and keep only the bosons
in $X^{M},P^{M}$ . In this gauge we have $g\left( \Theta \right) $ as given
in Eq.(\ref{gfixed}), where $\bar{\Theta}=\Theta ^{\dagger }\gamma ^{0}.$
The fermion $\Theta $ is classified as $\left( 4,\bar{4}\right) $ under SU$%
\left( 2,2\right) \times $SU$\left( 4\right) $ = SO$\left( 4,2\right) \times 
$SO$\left( 6\right) $ has 16 complex or 32 real fermionic components$.$The
action and the conserved charges $J=g^{-1}Lg$ are now expressed only in
terms of the 12D superspace variables $\left( X,P,\Theta \right) .$

The action, with this form of $g$ still has local kappa symmetry that was
embedded SU$\left( 2,2|4\right) _{L}$, but now the naive kappa
transformation must be followed by a $\Theta $-dependent SO$\left(
4,2\right) \times $SO$\left( 6\right) $ gauge transformation to keep the
gauge fixed form of $g$ unchanged. Similarly, there still is a global SU$%
\left( 2,2|4\right) _{R}$ symmetry, but this transformation must be
compensated by a local SO$\left( 4,2\right) \times $SO$\left( 6\right) $
gauge transformation. Therefore now both kappa and global supersymmetry
transformations also act on $\left( X,P\right) $ via the $\Theta $-dependent
SO$\left( 4,2\right) \times $SO$\left( 6\right) $ compensating gauge
transformations.

There is another way of fixing the local SO$\left( 4,2\right) \times $SO$%
\left( 6\right) $ symmetry. We can completely gauge away $\left( X,P\right) $
and shift all degrees of freedom to $g\left( \tau \right) .$ Then the first
term in the action can be completely eliminated by satisfying the
constraints $X^{2}=P^{2}=X\cdot P=0.$ The action reduces to only the second
term with $L$ a fixed matrix oriented in a particular direction in SO$\left(
4,2\right) \times $SO$\left( 6\right) $ space. With an appropriate
parametrization of $g$ this form of the action can be written in terms of
twistors by using the techniques of the third reference in \cite{super2t}.
Hence 12D superspace $\left( X,P,\Theta \right) $ can be converted to
supertwistor space and vice versa.

\subsection{Kappa gauge fixing}

Half of the fermions in $\Theta $ can be eliminated by choosing the kappa
gauge $\Gamma ^{+^{\prime }}\Theta =0,$ i.e. 
\begin{equation}
\Theta =\left( 
\begin{array}{c}
\theta ^{\alpha a} \\ 
0%
\end{array}%
\right) ,\quad \bar{\Theta}=\left( 
\begin{array}{cc}
0 & \bar{\theta}_{a}^{\dot{\alpha}}%
\end{array}%
\right) .
\end{equation}%
Here $a=1,\cdots ,4$ is an SU$\left( 4\right) $ index for the $\bar{4}$
(raised) or $4$ (lowered) representations, and $\alpha =1,2$ or $\dot{\alpha}%
=1,2$ are SL$\left( 2,C\right) =$SO$\left( 3,1\right) $ spinor indices. The $%
\theta ^{\alpha a},\bar{\theta}_{a}^{\dot{\alpha}},$ are related to each
other by Hermitian conjugation and multiplication by the SL$\left(
2,C\right) $ invariant Levi-Civita tensors $\varepsilon ^{\dot{\alpha}\dot{%
\beta}}$ or $\varepsilon _{\varepsilon \beta }$. This gauge fixing breaks
the manifest SO$\left( 4,2\right) \times $SO$\left( 6\right) $
classification of the fermions. Then $g$ contains only 8 complex fermions or
16 real fermions that are non-linearly coupled to the orbital AdS$_{5}\times 
$S$^{5}$ operators $L^{mn},L^{IJ}$ (which themselves are also non-linearly
realized as given in (\ref{lmnclass1}-\ref{lmnclass5})). The expansion of
the exponential yields the following form for $g$%
\begin{equation}
g=\left( 
\begin{array}{ccc}
\mathbf{1}_{2} & \frac{1}{2}\theta \bar{\theta} & \theta \\ 
0 & \mathbf{1}_{2} & 0 \\ 
0 & \bar{\theta} & \mathbf{1}_{4}%
\end{array}%
\right)  \label{fixedg}
\end{equation}%
With this form of $g$ both the action and the global SU$\left( 2,2|4\right)
_{R}$ charges $J=g^{-1}Lg$ simplify dramatically.

\subsection{Sp$\left( 2,R\right) $ gauge fixing and AdS$_{5}\times $S$^{5}$
versus M$^{4}\times $R$^{6}$}

The AdS$_{5}\times $S$^{5}$ curved background is created from the flat
12-dimensional background in a special Sp$\left( 2,R\right) $ gauge. Thus,
using the local Sp$\left( 2,R\right) $ symmetry we choose two gauges: the
component $P^{+^{\prime }}=\left( P^{0^{\prime }}+P^{1^{\prime }}\right) /%
\sqrt{2}$ of the SO$\left( 4,2\right) $ vector vanishes for all $\tau ,$ $%
P^{+^{\prime }}\left( \tau \right) =0,$ and the magnitude of the SO$\left(
6\right) $ vector\ $X^{I}\left( \tau \right) $ is taken as a $\tau $
independent constant $\left\vert X^{I}\left( \tau \right) \right\vert =R$ .
\ The remaining components of the 12 dimensional vectors $X_{i}^{M}=\left(
X^{M},P^{M}\right) $ are then parameterized as follows after solving
explicitly the two 12-dimensional constraints $X^{2}=X\cdot P=0$ 
\begin{align}
M& =\left( \,\,+^{\prime }\quad \quad \,\,\,-^{\prime }\quad \quad \,\quad
\,\mu \quad \quad \,\quad I\right)  \notag \\
X^{M}& =\frac{R}{\left\vert \mathbf{y}\right\vert }\left( R,\,\,\,\,\,\,\,%
\frac{x^{2}+\mathbf{y}^{2}}{2R},\quad x^{\mu }\,,\quad \mathbf{y}^{I}\right)
\label{adsX} \\
P^{M}& =\frac{\left\vert \mathbf{y}\right\vert }{R}\left( 0,\,\,\,\,\frac{1}{%
R}\left( x\cdot p+\mathbf{y\cdot k}\right) \,,\,\,\,\,\,\,\,\,\,p^{\mu
}\,\,\,,\,\,\mathbf{\,\,\,k}^{I}\right) .  \label{adsP}
\end{align}%
Here the $I$ are SO$\left( 6\right) $ indices as before, the $\mu $ are SO$%
\left( 3,1\right) $ indices, and the $m=\left( +^{\prime },-^{\prime },\mu
\right) $ are SO$\left( 4,2\right) $ indices in a lightcone type basis in
the extra dimensions $m=(0^{\prime },1^{\prime })\leftrightarrow \left(
+^{\prime },-^{\prime }\right) $. In this gauge the 12-dimensional flat
(10,2) metric becomes the 10 dimensional AdS$_{5}\times $S$^{5}$ metric
after expressing $\left( \mathbf{y}^{I}\mathbf{,k}^{I}\right) $ in terms of
radial and angular variables (see Appendix-B) 
\begin{equation}
ds^{2}=dX^{M}dX_{M}=\frac{R^{2}}{y^{2}}\left[ \left( dx^{\mu }\right)
^{2}+\left( dy\right) ^{2}\right] +\left( d\mathbf{\Omega }\right) ^{2}.
\label{metric}
\end{equation}%
The boundary of the AdS$_{5}$ space at $y\rightarrow 0$ is Minkowski space $%
x^{\mu }$ in 4-dimensions.

In the following we work in units of $R=1$ to simplify our expressions.
Inserting the gauge fixed form of $X_{i}^{M}$ into the Lagrangian (\ref{LL}%
), and using the definitions of radial and angular variables given in
Appendix-B at the classical level, gives 
\begin{equation}
\mathcal{L}=p\cdot \dot{x}+\mathbf{k}\cdot \mathbf{\dot{y}}-\frac{1}{2}%
A^{22}\left( \left( p^{2}+k^{2}\right) y^{2}+\frac{1}{2}L^{IJ}L_{IJ})\right)
-\frac{1}{16}Str\left( L\left( \partial _{\tau }gg^{-1}\right) \right) .
\label{adsaction11}
\end{equation}%
For the gauged fixed $g$ in Eq.(\ref{fixedg}) the second term simplifies to%
\begin{equation}
-\frac{1}{16}Str\left( L\left( \partial _{\tau }gg^{-1}\right) \right) =%
\frac{1}{2}\left( \partial _{\tau }\bar{\theta}_{a}\bar{\sigma}_{\mu }\theta
^{a}-\bar{\theta}_{a}\bar{\sigma}_{\mu }\partial _{\tau }\theta ^{a}\right)
\,p^{\mu }  \label{fermi}
\end{equation}

The term multiplying $A^{22}$ in Eq.(\ref{adsaction11}), after quantum
ordering, $P^{2}=y\left( p^{2}+k^{2}\right) y+\frac{1}{2}L^{IJ}L_{IJ}\sim $%
0, is the remaining 12-dimensional constraint. In Appendix-B we show in
detail that it has the interpretation of the Laplacian on AdS$_{5}\times $S$%
^{5}$ when applied on physical states 
\begin{equation}
P^{2}\phi =-\hbar ^{2}\nabla _{\left( AdS_{5}\times S^{5}\right) }^{2}\phi
\sim 0.
\end{equation}%
Another way of seeing this is to use the relations in Eqs.(\ref{c42}-\ref{c6}%
) that followed from the Sp$\left( 2,R\right) $ constraints, including $%
P^{2}=0,$ 
\begin{equation}
\left[ -C_{2}\left( 4,2\right) +C_{2}\left( 6\right) \right] \phi =0.
\end{equation}%
These too reflect the underlying 12 dimensions in the 2T formalism. Indeed,
we have $-\hbar ^{2}\nabla _{S_{5}}^{2}=C_{2}\left( 6\right) =\hbar
^{2}l\left( l+4\right) ,$ and if we compute $C_{2}\left( 4,2\right) =\frac{1%
}{2}L_{mn}L^{mn}=\frac{1}{2}L_{\mu \nu }L^{\mu \nu }-\left( L^{+^{\prime
}-^{\prime }}\right) ^{2}-L^{+^{\prime }\mu }L_{\,\,\,\mu }^{-^{\prime
}}-L_{\,\,\,\mu }^{-^{\prime }}L^{+^{\prime }\mu }$ by inserting an
appropriately quantum ordered version (see third reference in \cite{old2T})
of the non-linear $L_{mn}$ in Eqs.(\ref{lmnclass1}-\ref{lmnclass5}), we find
precisely $C_{2}\left( 4,2\right) \rightarrow +\hbar ^{2}\nabla
_{AdS_{5}}^{2}.$

It is worth noting another 1T-physics gauge choice, namely the relativistic
particle gauge given by $X^{+^{\prime }}=1$ and $P^{+^{\prime }}=0$ used in
many previous applications of 2T-physics. Again, solving explicitly the
constraints $X^{2}=X\cdot P=0$ the 12 dimensional vectors $X_{i}^{M}=\left(
X^{M},P^{M}\right) $ are parameterized as follows%
\begin{align}
M& =\left( \,\,+^{\prime }\quad \quad \,\,\,-^{\prime }\quad \quad \,\quad
\,\mu \quad \quad \,\quad I\right)  \notag \\
X^{M}& =\left( 1,\,\,\,\,\,\,\,\frac{1}{2}\left( x^{2}+\mathbf{x}^{2}\right)
,\quad x^{\mu }\,,\quad \mathbf{x}^{I}\right)  \label{mX} \\
P^{M}& =\left( 0,\,\,\,\,\left( x\cdot p+\mathbf{x\cdot p}\right)
\,,\,\,\,\,\,\,\,\,\,p^{\mu }\,\,\,,\,\,\mathbf{\,\,\,p}^{I}\right) .
\label{mP}
\end{align}%
where the surviving 10 coordinates $\left( x^{\mu }\,,\quad \mathbf{x}%
^{I}\right) $ now are in flat 10-dimensional Minkowski space, however the
interactions with the 16 $\theta ^{\prime }s$ break this space to $%
M^{4}\times R^{6}$ with linearly realized SO$\left( 3,1\right) \times $SO$%
\left( 6\right) $ symmetry. The gauge fixed action looks just like Eq.(\ref%
{adsaction11}) except for replacing $\left( \mathbf{y}^{I}\mathbf{,k}%
^{I}\right) $ by $\left( \mathbf{x}^{I}\mathbf{,p}^{I}\right) $ and
modifying the coefficient of $A^{22}$ to the form $-\frac{1}{2}A^{22}\left(
p^{2}+\mathbf{p}^{2}\right) .$

The two 1T-physics superparticle actions obtained by the above gauge fixing
represent the same 2T-physics system, but with different 1T Hamiltonians. In
this sense they are duals. They both have the same gauge invariant global
symmetry SU$\left( 2,2|4\right) $ realized in the same representations with
vanishing Casimirs as in Eq.(\ref{Cn}) at the quantum level.

Therefore we see signs of a possible duality between AdS$_{5}\times $S$^{5}$
supergravity and $M^{4}\times R^{6}$ supergravity which must be true at the
level of the kinetic term, but remains to be examined in the presence of
interactions.

\subsection{Global SU$\left( 2,2|4\right) _{R}$ supersymmetry generators}

Using the equations of motion that follow from the gauge fixed Lagrangian,
one may easily verify that there is a conserved SU$\left( 2,2|4\right) _{R}$
current that is none other than the gauge fixed form of the global current
given in (\ref{J}), $J=g^{-1}Lg.$ This is expected since the global symmetry
SU$\left( 2,2|4\right) _{R}$ in the original action, and the corresponding
symmetry current $J,$ commute with all the local symmetries Sp$\left(
2,R\right) ,$ SO$\left( 4,2\right) \times $SO$\left( 6\right) $ and kappa.
Any fixing of the gauge symmetries cannot destroy the physical global
symmetry. However, the transformation rules of the surviving degrees of
freedom become complicated because the naive SU$\left( 2,2|4\right) _{R}$
transformation of the gauge fixed $g$ must be compensated by a local $%
(x,p,\theta $-dependent$)$ kappa, SO$\left( 4,2\right) \times $SO$\left(
6\right) $ and Sp$\left( 2,R\right) $ gauge transformations in order to
maintain the form of the gauge fixed $g,X,P.$

The complicated nonlinear SU$\left( 2,2|4\right) _{R}$ transformations are
all taken into account automatically in the gauge fixed expression of the
global SU$\left( 2,2|4\right) _{R}$ charges $J=g^{-1}Lg.$ To generate the
correct transformations we only need to write these charges in terms of the
canonical variables of the surviving degrees of freedom and commute them
with any quantity that needs to be transformed. At the classical level
orders of factors of canonical variables are neglected.

To construct the gauge fixed charges $J$ we begin with the orbital $L^{MN}.$
The non-linear SO$\left( 10,2\right) $ generators are constructed in the AdS$%
_{5}\times $S$^{5}$ background by inserting the gauge fixed form of $\left(
X^{M},P^{M}\right) $ in $L^{MN}=X^{M}P^{N}-X^{N}P^{M}.$ We get the classical
expressions (not watching orders of factors\footnote{%
These are the classical expressions needed in the classical transformation
rules. Correct quantum ordering is not needed in the classical action in
this paper, but is needed in other contexts. See the third reference [1]
(hep-th/9810025) for the correct quantum ordering and verification of the
full SO$\left( 10,2\right) $ symmetry of the quantum ordered Laplacian.
Compared to this reference we have redefined $\mathbf{y}^{I}=\frac{\mathbf{u}%
^{I}}{\mathbf{u}^{2}}.$}) 
\begin{align}
L^{IJ}& =\mathbf{y}^{I}\mathbf{k}^{J}-\mathbf{y}^{J}\mathbf{k}^{I},\quad
L^{\mu \nu }=x^{\mu }p^{\nu }-x^{\nu }p^{\mu },  \label{lmnclass1} \\
L^{+^{\prime }-^{\prime }}& =\mathbf{y\cdot k}+x\cdot p,\quad L^{+^{\prime
}\mu }=p^{\mu },\quad L^{+^{\prime }I}=\mathbf{k}^{I},\quad \\
L^{-^{\prime }\mu }& =\frac{1}{2}\left( x^{2}+\mathbf{y}^{2}\right) p^{\mu
}-\left( x\cdot p+\mathbf{y\cdot k}\right) x^{\mu },  \label{lmnclass} \\
L^{-^{\prime }I}& =\frac{1}{2}\left( x^{2}+\mathbf{y}^{2}\right) \mathbf{k}%
^{I}-\left( x\cdot p+\mathbf{y\cdot k}\right) \mathbf{y}^{I}, \\
L^{\mu I}& =x^{\mu }\mathbf{\,k}^{I}-p^{\mu }\mathbf{y}^{I}.
\label{lmnclass5}
\end{align}%
These are easily expressed in terms of radial/angular variables as in
Appendix-B. The components of the SO$\left( 10,2\right) $ angular momenta $%
L^{IJ},$ $L^{mn},$ $L^{mI}$ that appear in the last term of the gauge fixed
Lagrangian in (\ref{adsaction11}), the gauge fixed current $J$ in (\ref{J}),
and the gauge fixed kappa transformation rules in (\ref{xi},\ref{delkappa})
now contain the non-linear forms given above.

For the $M^{4}\times R^{6}$ gauge the expressions for the $L^{MN}$ are
identical to those of AdS$_{5}\times $S$^{5}$ expressions above at the
classical level (except for replacing $\left( \mathbf{y}^{I}\mathbf{,k}%
^{I}\right) $ by $\left( \mathbf{x}^{I}\mathbf{,p}^{I}\right) $), but at the
quantum level the ordering of operators is different, and is probably the
same as flat 10D space (see the first reference in \cite{old2T}).

The various components of the 8$\times 8$ matrix $J$ can be identified as
the superconformal charges $P^{\mu },J^{\mu \nu },D,K^{\mu },Q,S$ and
R-symmetry charges $J^{IJ}$ as follows 
\begin{equation}
J=\left( 
\begin{array}{ccc}
J_{\,\beta }^{\alpha }+\frac{1}{2}\delta _{\,\beta }^{\alpha }D & K^{\alpha 
\dot{\beta}} & \bar{S}^{\alpha b} \\ 
P_{\dot{\alpha}\beta } & J_{\dot{\alpha}}^{\,\dot{\beta}}-\frac{1}{2}\delta
_{\dot{\alpha}}^{\,\dot{\beta}}D & \bar{Q}_{\dot{\alpha}}^{\,b} \\ 
Q_{a\beta } & S_{a}^{\,\dot{\beta}} & J_{a}^{\,b}%
\end{array}%
\right) ,
\end{equation}%
where, after using the standard definitions that convert $SL\left(
2,C\right) $ and SU$\left( 4\right) $ spinor indices to spacetime indices, 
\begin{eqnarray}
P_{\dot{\alpha}\beta } &=&\left( \sigma _{\mu }\right) _{\dot{\alpha}\beta
}P^{\mu },\quad K^{\alpha \dot{\beta}}=\left( \bar{\sigma}_{\mu }\right)
^{\alpha \dot{\beta}}K^{\mu }, \\
J_{\,\beta }^{\alpha } &=&\frac{1}{2}\left( \sigma _{\mu \nu }\right)
_{\,\beta }^{\alpha }J^{\mu \nu },\quad J_{\dot{\alpha}}^{\,\dot{\beta}}=%
\frac{1}{2}\left( \bar{\sigma}_{\mu \nu }\right) _{\dot{\alpha}}^{\,\dot{%
\beta}}J^{\mu \nu },\quad \\
J_{a}^{\,b} &=&\left( \Gamma _{IJ}\right) _{a}^{\,b}J^{IJ},\quad \bar{Q}%
=Q^{\dagger }\left( i\sigma _{2}\right) ,\quad \bar{S}=S^{\dagger }\left(
i\sigma _{2}\right) ,
\end{eqnarray}%
the generators of SU$\left( 2,2|4\right) $ are given in terms of the
dynamical variables $\left( x^{\mu },y,\mathbf{\Omega }\right) $ and their
canonical conjugates $\left( p^{\mu },k,L^{IJ}\right) $ as follows 
\begin{align}
P^{\mu }& =p^{\mu },\quad Q_{a\alpha }=\bar{\theta}_{a}^{\dot{\beta}}\,p_{%
\dot{\beta}\alpha },\quad \\
\bar{S}^{\alpha a}& =x^{\alpha \dot{\beta}}\bar{Q}_{\dot{\beta}%
}^{\,a}+\theta ^{\alpha b}L_{b}^{\,a}+\theta ^{\alpha b}Q_{b\beta }\,\theta
^{\beta a}, \\
J^{IJ}& =L^{IJ}+\frac{1}{2}\left( \theta \Gamma ^{IJ}Q+\bar{Q}\Gamma ^{IJ}%
\bar{\theta}\right) ,\quad \\
J^{\mu \nu }& =L^{\mu \nu }+\frac{1}{2}\left( Q_{a}\sigma ^{\mu \nu }\theta
^{a}+\bar{\theta}_{a}\bar{\sigma}^{\mu \nu }\bar{Q}^{a}\right) ,\quad \\
D& =yk+x\cdot p+\frac{1}{2}\left( Q_{a}\theta ^{a}+\bar{\theta}^{a}\bar{Q}%
_{a}\right) \\
K^{\mu }& =L^{-\mu }+\frac{1}{2}x^{\alpha \dot{\beta}}\bar{Q}_{\dot{\beta}%
}^{\,a}\bar{\theta}_{a}^{\dot{\gamma}}\left( \bar{\sigma}^{\mu }\right) _{%
\dot{\gamma}\alpha }+\frac{1}{2}\left( \bar{\sigma}^{\mu }\right) _{\dot{%
\gamma}\alpha }\theta ^{\alpha a}Q_{a\beta }x^{\beta \dot{\gamma}}  \notag \\
& +\theta ^{\alpha b}L_{b}^{\,a}\bar{\theta}_{a}^{\dot{\gamma}}\left( \bar{%
\sigma}^{\mu }\right) _{\dot{\gamma}\alpha }+\frac{1}{4}\theta ^{\alpha
b}Q_{b\beta }\,\theta ^{\beta a}\bar{\theta}_{a}^{\dot{\gamma}}\left( \bar{%
\sigma}^{\mu }\right) _{\dot{\gamma}\alpha }
\end{align}%
where $L^{\mu \nu },L^{IJ},L^{-\mu }$ and $L_{a}^{\,b}=\left( \Gamma
_{IJ}\right) _{a}^{\,b}L^{IJ}$ are given in (\ref{lmnclass1}-\ref{lmnclass5}%
). The case for the $M^{4}\times R^{6}$ gauge is similar.

To find the non-linear action of the SU$\left( 2,2|4\right) _{R}$ symmetry
on the 10D super phase space variables we only need to commute the
generators given above with the remaining canonical variables, or any other
operator $\mathcal{O}$ constructed from the canonical variables 
\begin{equation}
\delta _{\varepsilon }\mathcal{O}=-i\left[ Str\left( \varepsilon J\right) ,%
\mathcal{O}\right] ,
\end{equation}%
where $\varepsilon $ is the infinitesimal global parameters of the original
SU$\left( 2,2|4\right) _{R}$. For details of similar discussions of
nonlinearly realized symmetries see \cite{old2T}. In the present problem we
also need to know the commutation rules for fermions. From Eq.(\ref{fermi})
we see that the canonical conjugate to $\bar{\theta}_{a}^{\alpha }$ is $\pi
_{a}^{\alpha }=\frac{1}{2}\left( \bar{\sigma}_{\mu }\theta ^{a}\right)
_{\alpha }p^{\mu }.$ This is a second class constraint which can be solved
(since $p^{2}\neq 0$), by taking the fundamental non-zero (anti)commutation
rules for fermions to be 
\begin{equation}
\left\{ \bar{\theta}_{a}^{a},\theta ^{\dot{\beta}b}\right\} =\hbar \frac{%
2p^{\mu }\left( \sigma _{\mu }\right) _{\alpha \dot{\beta}}}{p^{2}}\delta
_{a}^{b}.  \label{thetas}
\end{equation}%
After this, obtaining the $\delta _{\varepsilon }\mathcal{O}$ for any $%
\mathcal{O}$ is a straightforward computation. In particular, it can be
shown that the gauge fixed action is invariant under the nonlinear
transformations generated by these SU$\left( 2,2|4\right) $ charges, as
should be expected by construction.

Note that the gauge invariant constraints $\left( J\right)
^{2}=-4C_{2}\left( 6\right) \mathbf{1+}8i\hbar \left( J\right) $\textbf{\ }%
and vanishing Casimir conditions\textbf{\ }$C_{n}\left( 2,2|4\right) =0$ of
Eqs.(\ref{j2},\ref{Cn}) are automatically satisfied by the generators given
above (at either the classical level, or quantum level, after appropriate
quantum ordering in any gauge).

\subsection{Spectrum and the AdS$_{5}\times $S$^{5}$ supersymmetric
Kaluza-Klein towers}

The gauge fixed action describes the entire AdS$_{5}\times $S$^{5}$
supersymmetric Kaluza-Klein tower given in previous literature \cite{MG}\cite%
{KRN}. To see this, it is worth noticing the following simple observations.
First and foremost, the global symmetry of the theory is SU$\left(
2,2|4\right) _{R},$ therefore all states must fall into infinite dimensional
unitary representations of this group. Such representations are fully
characterized by their lowest states labelled by the compact bosonic
subgroup SU$\left( 2\right) _{1}\times $SU$\left( 2\right) _{2}\times $SU$%
\left( 4\right) $. The specific representations that are realized are built
from the canonical degrees of freedom given above. Once the graviton
Kaluza-Klein tower is determined, the SU$\left( 2,2|4\right) $ supersymmetry
guarantees that the full super multiplet structure is also present at all
Kaluza-Klein levels. Such simple facts are also the determining factor of
the representation content of supergravity solutions \cite{MG}, but here
they occur in the context of the superparticle. Hence the representations
are identical once the supergravity multiplet is determined.

The quantum physical states are represented by a superfield with 8 complex $%
\theta ^{\prime }$s and the position coordinates (actually there is also
dependence on $\bar{\theta}$ like in a chiral superfield). In the following
we deal with the AdS$_{5}\times $S$^{5}$ case (the $M^{4}\times R^{6}$ case
is similar and simpler). Thus, we consider the physical states in the form
of the superfield 
\begin{equation}
\Phi \left( x^{\mu },\mathbf{y}^{I},\theta \right) =\Phi \left( x^{\mu },y,%
\mathbf{\Omega }^{I},\theta \right) .
\end{equation}%
It is a simple exercise to identify the 2$^{8}$ fields corresponding to 128
bosons and $128$ fermions as the 10-dimensional AdS$_{5}\times $S$^{5}$
supergravity fields\footnote{%
This can be understood by going to the \textquotedblleft rest
frame\textquotedblright\ of $p^{\mu }$ in which spin is identified as the SO$%
\left( 3\right) $ subgroup. Then, with $\alpha \leftrightarrow \dot{\alpha}$
doublet indices identified as SO$\left( 3\right) =$SU$\left( 2\right) $ spin
indices, the 8 $\theta $'s behave like spin=1/2 creators and the 8 $\bar{%
\theta}$'s like annihilators, according to Eq.(\ref{thetas}). Applying all
possible powers of 8 fermionic creators on a vacuum gives 128 bosons and 128
fermions. The components of the superfield identified in this
\textquotedblleft rest\textquotedblright\ frame are in one-to-one
correspondence to the supergravity states. This is verified by the SO$\left(
3\right) $ reduction of the SU(2)$\times $SU$(2)$= SO$\left( 4\right) $ list
of states provided in \cite{zero}. There SO$\left( 3\right) \subset $ SU(2)$%
\times $SU$(2)\subset $ SU$\left( 2,2\right) \subset $ SU$\left(
2,2|4\right) $, while here the same SO$\left( 3\right) $ occurs in the chain
SO$\left( 3\right) \subset $ SO$\left( 3,1\right) \subset $ SU$\left(
2,2\right) \subset $ SU$\left( 2,2|4\right) ,$ which is is why these
supergravity fields, labelled differently, can be put into one to one
correspondence. Actually, the analytic continuation of SU(2)$\times $SU$(2)$%
=SO$\left( 4\right) $ to SL$\left( 2,C\right) =$ SO$\left( 3,1\right) $
allows the rewriting of the list in covariant SO$\left( 3,1\right) $
notation as outlined in \cite{zero}.}. We have already constructed the
generators of SU$\left( 2,2|4\right) $ acting on this basis, and therefore
shown that this superfield is a basis for a non-linear representation of SU$%
\left( 2,2|4\right) .$ Thus the supergravity multiplet is evident, while the
Kaluza-Klein towers are simply the expansion of the superfield in harmonics
of S$^{5}.$

This superfield is subject to the remaining constraint that corresponds to
the vanishing of the AdS$_{5}\times $S$^{5}$ Laplacian that arises as
follows. The physical states in our AdS$_{5}\times $S$^{5}$ gauge must be
consistent with the Sp$\left( 2,R\right) $ gauge invariance, implying that
they should satisfy the constraints $X_{i}\cdot X_{j}\sim 0$. Although the
constraints $X^{2}=X\cdot P=0$ have been explicitly solved in the chosen Sp$%
\left( 2,R\right) $ gauge, the remaining constraint $P^{2}=0$, or
equivalently the AdS$_{5}\times $S$^{5}$ Laplacian, takes the following form
on the graviton tower 
\begin{equation}
Casimir\,\,of\,\,SO\left( 4,2\right) -Casimir\,\,of\,\,SO\left( 6\right) =0.
\end{equation}%
This is the same as the AdS$_{5}\times $S$^{5}$ Laplacian as shown in
Appendix-B. Thus, the mass of the field as given by the Casimir\thinspace
\thinspace of\thinspace \thinspace SO$\left( 4,2\right) $ is now fully
determined by the Casimir\thinspace \thinspace of\thinspace \thinspace SO$%
\left( 6\right) .$ In turn, the Casimir\thinspace \thinspace of\thinspace
\thinspace SO$\left( 6\right) $ is determined by the SO$\left( 6\right) $
representations that can be constructed from the traceless symmetrized
products of the six $y^{I}$. The traceless tensor with $l$ indices has the SO%
$\left( 6\right) $ Casimir eigenvalue $l\left( l+4\right) .$ Thus, the mass
of the Kaluza-Klein modes of the graviton supermultiplet and all of its AdS$%
_{5}\times $S$^{5}$ partners are determined by the integers $l=0,1,2,\cdots
. $

For the SU$\left( 2,2|4\right) $ classification of the Kaluza-Klein towers,
the orbital part is expressed in angular momentum space SO$\left( 4,2\right)
\times $SO$\left( 6\right) $ instead of 10D configuration space $\left(
x^{\mu },\mathbf{y}^{I}\right) $. The total SO$\left( 6\right) =$SU$\left(
4\right) $ representation of each member of the SU$\left( 2,2|4\right) $
supermultiplet is then determined by combining the orbital SO$\left(
4,2\right) \times $SO$\left( 6\right) $ quantum numbers of the state with
those SO$\left( 4,2\right) \times $SO$\left( 6\right) $ quantum numbers
coming from the 128$_{B}+128_{F}$ multiplet. This was done in \cite{KRN}.
Such representations are more conveniently described in the oscillator
formalism \cite{barsgunaydin}\cite{MG}\cite{zero}.

\section{Discussion}

The 2T-physics approach to AdS$_{5}\times $S$^{5}$ superspace introduced a
natural underlying 12-dimensional superspace and its associated symmetries.
Evidence of the hidden 12-dimensional structure may be expected in
supergravity. In particular this approach predicts that the Casimir
operators of SU$\left( 2,2|4\right) ,$ i.e. Str$\left( J^{n}\right) $ for
any $n,$ applied on the superfield must vanish as a consequence of the 12D
constraints $X^{2}=P^{2}=\left( X\cdot P+P\cdot X\right) =0$ applied on
physical states. This is true either in the fully covariant 12D quantization
or the partially gauge fixed 10D quantization on AdS$_{5}\times $S$^{5}$ or $%
M^{4}\times R^{6}$ or others. Therefore, it must be true in the supergravity
spectrum.

The representations of SU$\left( 2,2|4\right) $ that arise here are highly
limited by the algebraic constraints on its generators $\left( J\right) $
written as an 8$\times 8$ supermatrix 
\begin{equation}
\left( J\right) ^{2}=\frac{\hbar ^{2}}{4}l\left( l+4\right) \mathbf{1-}%
2\hbar \left( J\right) ,  \label{JJ2}
\end{equation}%
as derived in Eq.(\ref{j2}). In particular the fields of supergravities on
AdS$_{5}\times $S$^{5}$ and M$^{4}\times $R$^{6}$ or other 2T-dual versions,
which are equivalent to the quantum states of our 12D superparticle, must be
consistent with these algebraic constraints. We conjecture that any SU$%
\left( 2,2|4\right) $ representation that satisfies these constraints must
be uniquely equivalent to the 12-dimensional superparticle representation
presented in this paper.

Indeed direct computation within AdS$_{5}\times $S$^{5}$ supergravity, using
the independent oscillator formalism, has already shown that all Casimir
operators do have a universal zero eigenvalue for all Kaluza-Klein towers 
\cite{zero}, consistent with Eq.(\ref{JJ2}). In further work on the
oscillator formalism \cite{oscillators} we have shown that the oscillator
states that satisfy the algebraic contraints above indeed form a unique
subset within the vast Fock space, such that they uniqely give the SU$\left(
2,2|4\right) $ representations that correspond to all the Kaluza-Klein
towers of AdS$_{5}\times $S$^{5}$ supergravity. Additional evidence of the
12D structure related to Eq.(\ref{JJ2}) is likely to emerge. In particular
the suggested duality between AdS$_{5}\times $S$^{5}$ and M$^{4}\times $R$%
^{6}$ supergravity would be an interesting avenue to explore. According to
the results of this paper this duality must exist at the level of the
spectra (or the kinetic terms) and it would be interesting to investigate it
including the supergravity interactions.

We remind the reader that there are multiple advantages of the 2T
formulation. First it displays explicitly the $d+2$ higher dimensional
spacetime supersymmetries that are hidden in 1T-physics. Second, by
different gauge choices, it provides a family of 1T-physics dynamical
systems (different Hamiltonians\footnote{%
The reason that there are many 1T theories under the umbrella of the same 2T
theory, is because there is a single worldline parameter $\tau $ but there
are two timelike dimensions in target spacetime $X^{M}\left( \tau \right) .$
The two times $X^{0^{\prime }}\left( \tau \right) ,X^{0}\left( \tau \right) $
are not introduced arbitrarily, rather their presence is required as a
consequence of the local Sp$\left( 2,R\right) $ symmetry. In the gauge
fixing process one must decide which combination of the two times will be
identified with $\tau .$ Each such choice corresponds to a different
rearrangement of the evolution of the 2T system (in terms of fewer 1T
degrees of freedom) and therefore results in a different Hamiltonian that
represents a 1T dynamical system. Thus, a rich set of dualities is
established among the members of a family of 1T dynamical systems through
the 2T formalism. These ideas were illustrated with simple examples in \cite%
{old2T}-\cite{survey2T}. The existence of previously unknown hidden
symmetries and duality relations among simple systems (such as free
particle, hydrogen atom, harmonic oscillator, particle in AdSxS backgrounds,
etc.) is part of the evidence for the validity and effectiveness of the
underlying 2T structure.} in $d$ dimensions) that are dual to each other in
the sense that they can be related to each other under the gauge
transformations in the enlarged space \cite{old2T}. These gauge
transformations become duality type transformations among the gauge fixed 1T
dynamical systems. Each member in the 1T family of dynamical systems in $d$
dimensions holographically represents the same 2T system in $d+2$
dimensions. We have not explored the multiple 1T systems in this paper,
except for the two cases AdS$_{5}\times $S$^{5}$ and $M^{4}\times R^{6}$ but
it would be very interesting to pursue this avenue in the future to find
multiple dual systems related to supergravity.

Are there generalizations of the superparticle action considered in this
paper? For cases, such as AdS$_{4}\times $S$^{7}$, AdS$_{7}\times $S$^{4},$
AdS$_{6}\times $S$^{2},$ etc., with symmetries OSp$\left( 8^{\ast }|4\right) 
$, $F_{4},$ etc., one is tempted to extend the superparticle actions based
on OSp$\left( 8^{\ast }|4\right) $ and $F_{4}$ in dimensions $d=3,5,6$
mentioned in the introduction, by adding additional coordinates $X^{I}$ and
coupling them to the supergroup element $g\left( \tau \right) $ as we did
for SU$\left( 2,2|4\right) $. However, the resulting theory does not seem to
describe the spectrum of supergravity compactified on AdS$_{7}\times S^{4}$
etc., but rather some other spectrum which is interesting anyway.

For example, consider the case of $G=$OSp$\left( 8^{\ast }|4\right) .$ The
action is the same as Eqs.(\ref{LL}) with $\left( 11,2\right) $ dimensions $%
\left( X^{M},P^{M}\right) ,$ and Eq.(\ref{Lmatrix}) is replaced by%
\begin{equation}
L=\left( 
\begin{array}{cc}
\frac{1}{2}L^{mn}\Gamma _{mn} & 0 \\ 
0 & -L^{IJ}\Gamma _{IJ}%
\end{array}%
\right)  \label{Lmatrix11}
\end{equation}%
The extra factor of $\frac{1}{2}$ in the first entry of the matrix $L$ is
required, due to the 8 dimensions of the spinor in $\left( 6,2\right) $
dimensions, to maintain the local SO$\left( 6,2\right) \times $SO$\left(
5\right) $ symmetry\footnote{%
Generally the full factor in front of $L^{mn}$ (similarly $L_{IJ}$),
including the 1/16 in the action of Eq.(\ref{LL}), is $\frac{1}{4s_{D}}%
\Gamma _{mn}L^{mn}$, where $s_{D}$ is the size of the spinor in D-dimensions.%
}. This action has global symmetry OSp$\left( 8^{\ast }|4\right) $ and local
symmetry Sp$\left( 2,R\right) \times $SO$\left( 6,2\right) \times $SO$\left(
5\right) ,$ but it fails to have the kappa supersymmetry. The reason is that
the extra factor of $\frac{1}{2}$ needed for local SO$\left( 6,2\right)
\times $SO$\left( 5\right) $ spoils the correct relative coefficient needed
for the automatic SO$\left( 11,2\right) $ in the steps that led to Eq.(\ref%
{correct}). Therefore, there is no local kappa symmetry, which means the
number of $\Theta ^{\prime }$s cannot be cut down from 32 real components to
16 real components. Therefore the spectrum is described by a superfield $%
\Phi \left( x^{\mu },y,\Omega ,\Theta \right) $ with 32 real $\Theta
^{\prime }$s. It has $2^{15}$ bosonic and 2$^{15}$ fermionic fields as
functions of AdS$_{7}\times S^{4}$ (or $M^{6}\times R^{5} $), with quantum
numbers that correspond to the first massive level of the 11D supermembrane 
\cite{membrane}. This supermultiplet is much larger than the supergravity
multiplet with 128 bosons and 128 fermions. This superfield may be
interesting for investigations in 11-dimensional $M$-theory, but is not
useful to investigate AdS$_{4}\times $S$^{7}$ or AdS$_{7}\times S^{4}$
supergravity.

There are supergroups and appropriate $d+2$ spacetime dimensions that would
parallel the structure of SU$\left( 2,2|4\right) $ action of this paper,
including the local kappa supersymmetry. The crucial point is the size of
the spinor representations for the dimensions associated with $X^{m}$ and $%
X^{I}.$ When the spinor is of the same size the kappa supersymmetry is
automatically present. A straightforward case is the supergroup SU$\left(
1,1|2\right) $ applied to AdS$_{2}\times $S$^{2}$ (or $R^{1}\times R^{3})$
as exact analog of the present paper, but in 4+2 dimensions instead of 10+2
dimensions. Additional cases include

\begin{itemize}
\item The supergroup SU$\left( 1,1|2\right) \times $SU$\left( 1,1|2\right) $
with $X^{M}=$ $\left( 10,2\right) $ such that $X^{m}=\left( 2,1\right)
+\left( 2,1\right) $ and $X^{I}=\left( 3,0\right) +\left( 3,0\right) $ are
coupled via the $L^{MN}$ to the four bosonic subgroups. After eliminating
all the bosons from $g\left( \tau \right) ,$ the remaining super phase space
has $\left( X^{M},P^{M},\Theta \right) $ with 8 complex or 16 real $\Theta $%
's. The kappa supersymmetry cuts down the fermions to 8 real components and
Sp$\left( 2\right) $ reduces the bosonic space to 10 dimensions with one
time. Therefore the physical states are described by a superfield with $8$
bosons and 8 fermions in 10 dimensions. This spectrum must correspond to a
compactified version of 10-dimensional super Yang Mills theory with a global
symmetry of SU$\left( 1,1|2\right) \times $SU$\left( 1,1|2\right) .$

\item The supergroup SU$\left( 1,1|2\right) \times $SU$\left( 2|2\right) $
with $X^{M}=$ $\left( 10,2\right) $ and $X^{m}=\left( 2,1\right) $ and $%
X^{I}=\left( 3,0\right) +\left( 3,0\right) +\left( 3,0\right) $ coupled via
the $L^{MN}$ to the bosonic subgroups. This must correspond again to another
form of compactified super Yang Mills theory with a global symmetry of SU$%
\left( 1,1|2\right) \times $SU$\left( 2|2\right) .$
\end{itemize}

Other supergroups provide generalizations in a different direction, by
having additional bosonic coordinates in $g\left( \tau \right) $ that cannot
be gauged away. In appropriate cases the extra bosonic coordinates can be
interpreted precisely as collective coordinates (providing charges)\ for
D-branes. In particular the supergroup $G=$OSp$\left( 1|64\right) $ gives an
interesting \textquotedblleft Toy M-model\textquotedblright\ described
briefly in previous publications \cite{super2t}.

A similar situation can arise with alternative schemes in constructing
models using OSp$\left( 8|4\right) $ (or $F_{4}$, etc.). As mentioned above,
if we insist on SO$\left( 6,2\right) \times $SO$\left( 5\right) $ (for $%
F_{4} $: SO$\left( 5,2\right) \times $SO$\left( 3\right) $) local symmetry
to gauge away all bosons in $g\left( \tau \right) $, then we cannot have
kappa supersymmetry because the spinor dimensions of these subgroups are
different from one another. However, there is another alternative: we can
insist on kappa supersymmetry, which is guaranteed for $g\left( \tau \right)
\in $OSp$\left( 8|4\right) $ (or $F_{4}$, etc.), if we take the same
coefficient in the blocks of $L$ (i.e. when $L$ has the form of Eq(\ref%
{Lmatrix}) instead of Eq.(\ref{Lmatrix11})). Then, with the factor 1/32 in
the second term of the action Eq(\ref{LL}), there is local SO$\left(
6,2\right) $ symmetry (or SO$(5,2)$ for F$_{4}$), but only global SO$\left(
5\right) $ (for $F_{4}$: SO$\left( 3\right) $) symmetry\footnote{%
Similarly, for the AdS$_{4}\times $S$^{7}$ model, with the analytically
continued $g\left( \tau \right) \in $OS$\left( 8,4\right) ,$ we take the
factor 1/16 to have local SO$\left( 3,2\right) $ and global SO$\left(
8\right) $ symmetry. Also, for AdS$_{2}\times $S$^{6}$ model, with the
analytically continued $g\left( \tau \right) \in F_4 ,$ we take the factor
1/8 to have local SO$\left( 1,2\right) $ and global SO$\left( 7\right) $
symmetry.}. This implies that the Sp$\left( 4\right) $ (for $F_{4} $: SU$%
\left( 2\right) $) bosons in $g\left( \tau \right) $ remain as additional
bosonic degrees of freedom. The extra bosons correspond to $D$-brane
collective degrees of freedom. Of course, now there are half as many
physical $\Theta $'s thanks to the local kappa symmetry. With 16 remaining
physical fermions (8 for $F_{4}$), one obtains the 11D supergravity states
(10D Yang-Mills states for $F_{4}$). Therefore the quantum states of the
model must correspond to the AdS$_{7}\times $S$^{5}$ compactified version of
11-dimensional supergravity in the presence of D-branes, with nontrivial
charges in the 5-dimensions (for $F_{4}:$ AdS$_{6}\times $S$^{2}\times $S$%
^{2}$ compactification of 10D super Yang Mills).

For all the generalized cases of the type described in the previous
paragraph, the global supersymmetry $G$ has charges $\left( J\right) $
written in the form of a supermatrix $J=\frac{i}{4}g^{-1}Lg.$ By the same
arguments as SU$\left( 2,2|4\right) $ that we discussed in Eqs.(\ref{xixj}-%
\ref{j2}), we obtain the form 
\begin{equation}
\left( J\right) ^{2}=\mathbf{-}2\hbar \left( J\right) +\frac{\hbar ^{2}}{4}%
l\left( l+n-2\right) \mathbf{1-}\frac{\hbar ^{2}}{16}\left(
d^{2}-n^{2}\right) g^{-1}\left( 
\begin{array}{cc}
1 & 0 \\ 
0 & 0%
\end{array}%
\right) g  \label{JJ22}
\end{equation}%
where $l=0,1,2,\cdots $ is the angular momentum of the harmonics on $S^{n}$
as discussed in Appendix-B, and $d,n$ correspond to the AdS$_{d}\times $S$%
^{n}$ described by the model. The last term arises from reordering operators
in Eqs.(\ref{c42}-\ref{c6}) before the constraints in Eq.(\ref{xixj2}) can
be applied. This equation is an algebraic constraint on the representations
of $G,$ which distinguishes the ones relevant for the superparticle
described by the corresponding action. It determines all properties of the
representation, including the physical states, their representation
properties, and eigenvalues of all the Casimir operators. In particular,
since the supertace of the right hand side does not vanish, these Casimir
eigenvalues do not vanish. They are predicted to be certain numbers that
depend on the representation labels, i.e. 
\begin{equation}
C_{2}=\frac{1}{2}Str\left( J^{2}\right) =\frac{\hbar ^{2}}{8}l\left(
l+n-2\right) \text{Str}\left( \mathbf{1}\right) -\frac{\hbar ^{2}}{16}\left(
d^{2}-n^{2}\right) \text{Str}\left( 
\begin{array}{cc}
1 & 0 \\ 
0 & 0%
\end{array}%
\right) ,
\end{equation}%
while higher Casimirs are computed by repeatedly applying Eq.(\ref{JJ22}).
Thus, these 2T-physics models make rather strong predictions which can be
tested by analyzing the representation content of the corresponding field
theories. The verification of such predictions would imply the existence of
the hidden higher dimensional structures in the corresponding field theories.

These and other generalizations of interest will be further studied in
future publications.

\appendix

\subsection{Appendix-A (gamma matrices)}

The gamma matrices for SO$\left( 4,2\right) =SU\left( 2,2\right) $ connect
the $4$ and $\bar{4}$ spinor representations. They may be taken as $\Gamma
_{m}=\left( 1,\gamma _{0},\gamma _{1},\gamma _{2},\gamma _{3},\gamma
_{5}\right) $ for $4\times \bar{4}$ or $\bar{\Gamma}_{m}=\left( -1,\gamma
_{0},\gamma _{1},\gamma _{2},\gamma _{3},\gamma _{5}\right) $ for $\bar{4}%
\times 4,$ which differ in the representation of $\Gamma _{0^{\prime }}=\pm
1 $ respectively, and satisfy $\Gamma _{m}\bar{\Gamma}_{n}+\Gamma _{n}\bar{%
\Gamma}_{m}=2\eta _{mn}.$ The $\gamma _{\mu },\gamma _{5}$ are the standard
four dimensional Dirac gamma matrices$.$ For convenience we may take $\gamma
_{0}$ antihermitian, the others Hermitian. The 15 traceless SU$\left(
2,2\right) $ generators in the $4$ representation $\frac{1}{2}\Gamma _{mn}$
are given by $\Gamma _{mn}=\frac{1}{2}\left( \Gamma _{m}\bar{\Gamma}%
_{n}-\Gamma _{n}\bar{\Gamma}_{m}\right) ,$ i.e. $\Gamma _{0^{\prime
}5}=\gamma _{5},$ $\Gamma _{0^{\prime }\mu }=\gamma _{\mu },$ $\Gamma _{5\mu
}=\gamma _{5}\gamma _{\mu }$, $\Gamma _{\mu \nu }=\left[ \gamma _{\mu
},\gamma _{\nu }\right] /2.$ These satisfy $\Gamma _{0}\left( \Gamma
_{m}\right) ^{\dagger }\Gamma _{0}=\bar{\Gamma}_{m}$ and $\Gamma _{0}\left(
\Gamma _{mn}\right) ^{\dagger }\Gamma _{0}=\Gamma _{mn}.$ Similarly one
constructs the fifteen SO$\left( 6\right) =$SU$\left( 4\right) $ generators
by using the analytic continuation of the two timelike directions that give $%
\Gamma _{I}=\left( i,\gamma _{4},\gamma _{1},\gamma _{2},\gamma _{3},\gamma
_{5}\right) $ or $\bar{\Gamma}_{I}=\left( -i,\gamma _{4},\gamma _{1},\gamma
_{2},\gamma _{3},\gamma _{5}\right) ,$ with $\gamma _{4}=i\gamma _{0}.$ In
this case $\bar{\Gamma}_{I}=$ $\Gamma _{I}^{\dagger }$ and $\left( \Gamma
_{IJ}\right) ^{\dagger }=-\Gamma _{IJ}.$ In this notation the inverse of the
SU$\left( 2,2|4\right) $ group element $g$ is given by $g^{-1}=C^{-1}g^{%
\dagger }C,$ where $C=\left( 
\begin{array}{cc}
\Gamma _{0} & 0 \\ 
0 & 1%
\end{array}%
\right) .$

Another basis for SO$\left( 4,2\right) $ gamma matrices is the one used in 
\cite{super2t}. We use $\Gamma ^{\pm ^{\prime }}=\pm \tau ^{\pm }\times 1,$ $%
\Gamma ^{0}=\tau _{3}\times i\sigma _{2},$ $\Gamma ^{1}=\tau _{3}\times
\sigma _{1},$ $\Gamma ^{3}=\tau _{3}\times \sigma _{3},$ which are purely
real, and the last one proportional to the identity which is purely
imaginary $\Gamma ^{2}=i1\times 1.$ Then define $\Gamma ^{m}=\left( \Gamma
^{\pm ^{\prime }},\Gamma ^{0},\Gamma ^{1},\Gamma ^{2},\Gamma ^{3}\right) $
and $\bar{\Gamma}^{m}=\left( \Gamma ^{\pm ^{\prime }},\Gamma ^{0},\Gamma
^{1},-\Gamma ^{2},\Gamma ^{3}\right) .$ These satisfy $\Gamma ^{m}\bar{\Gamma%
}^{n}+\Gamma ^{n}\bar{\Gamma}^{m}=2\eta ^{mn}.$ The SU$\left( 2,2\right) $
generators $\frac{1}{2}\Gamma _{mn}$ are given in terms of $\Gamma _{mn}=%
\frac{1}{2}\left( \Gamma _{m}\bar{\Gamma}_{n}-\Gamma _{n}\bar{\Gamma}%
_{m}\right) .$ These satisfy $u^{-1}\left( \Gamma ^{m}\right) ^{\dagger }u=%
\bar{\Gamma}^{m}$ and $u^{-1}\left( \Gamma ^{mn}\right) ^{\dagger }u=-\Gamma
^{mn},$ where $u=\tau _{1}\times i\sigma _{2}$ is the charge conjugation
matrix. Similarly, define SO$\left( 6\right) $ gamma matrices $\Gamma ^{I}$
by analytic continuation of the above, $\Gamma ^{6}=i\Gamma ^{0^{\prime }}$
and $\Gamma ^{4}=i\Gamma ^{0}.$ Then $\left( \Gamma ^{I}\right) ^{\dagger }=%
\bar{\Gamma}^{I}$ (i.e. $\Gamma ^{2}$ changes sign) and the generators are
antihermitian $\left( \Gamma ^{IJ}\right) ^{\dagger }=-\Gamma ^{IJ}.$ Then
the group element satisfies $g^{-1}=C^{-1}g^{\dagger }C$ where $C=\left( 
\begin{array}{cc}
u & 0 \\ 
0 & 1%
\end{array}%
\right) .$

\subsection{Appendix-B (12D constraints and Laplacian on AdS$_{5}\times $S$%
^{5}$)}

In this appendix we relate the 12-dimensional constraint $P^{2}=0$ to the
Laplacian on AdS$_{5}\times $S$^{5}$ 
\begin{equation}
\nabla _{\left( AdS_{5}\times S^{5}\right) }^{2}\sim 0.
\end{equation}%
We have seen that the AdS$_{5}\times $S$^{5}$ gauge in Eqs.(\ref{adsX},\ref%
{adsP}) already solves the 12D constraints $X^{2}=X\cdot P=0.$ After quantum
ordering, the remaining constraint takes the form $P^{2}=y\left( p^{2}+%
\mathbf{\,k}^{2}\right) y,$ where $y=\left\vert \mathbf{y}\right\vert $ is
the radial component of the six dimensional vector $\mathbf{y}^{I}.$ It will
be handled quantum mechanically by applying it on physical states $\Psi
\left( x^{\mu },\mathbf{y}\right) $ 
\begin{equation}
P^{2}\Psi \left( x^{\mu },\mathbf{y}\right) =y\left( p^{2}+\mathbf{\,k}%
^{2}\right) y\,\Psi \left( x^{\mu },\mathbf{y}\right) =0.  \label{phys}
\end{equation}

To proceed, we define the six Cartesian variables $\mathbf{y}^{I}$ in terms
of angular and radial variables $\mathbf{y}^{I}\mathbf{=}y\mathbf{\Omega }%
^{I}$, where $\mathbf{\Omega }$ \ is a unit vector that describes S$^{5}$, $%
\mathbf{\Omega }^{2}=1,$ and $y$ is the radial variable. The corresponding
momenta are then 
\begin{equation}
\mathbf{k}^{I}\mathbf{=}k\mathbf{\Omega }^{I}-\frac{1}{2y}\left( L^{IJ}%
\mathbf{\Omega }^{J}+\mathbf{\Omega }^{J}L^{IJ}\right) ,
\end{equation}%
where $\left( y,k\right) $ are canonical radial variables defined by $%
y=\left\vert \mathbf{y}\right\vert $ and $k=\left( \mathbf{k\cdot \Omega
+\Omega \cdot k}\right) /2$. Evidently the radial variables $\left(
y,k\right) $ commute with the SO$\left( 6\right) $ generators $L^{IJ}.$ The
SO(6) generators $L^{IJ}=\mathbf{y}^{I}\mathbf{k}^{J}-\mathbf{y}^{J}\mathbf{k%
}^{I}$ are expressed purely in terms of angular variables and their
derivatives on $S^{5},$ while the radial variables $\left( y,k\right) $
together with the 4-dimensional Minkowski variables $\left( x^{\mu },p^{\mu
}\right) $ make up the 5 canonical pairs on AdS$_{5}.$ The canonical
operators in this appendix are quantum ordered. In particular, one computes 
\begin{equation}
\mathbf{k}^{2}=k^{2}+\frac{1}{y^{2}}(\frac{1}{2}L^{IJ}L_{IJ}+\frac{\left(
n-1\right) \left( n-3\right) }{4}\hbar ^{2}).
\end{equation}%
where we wrote the quantum ordering term more generally as $\left(
n-1\right) \left( n-3\right) \hbar ^{2}/4$ in $n$ dimensions, but here we
need $n=6,$ which yields $\frac{15}{4}\hbar ^{2}.$ In the classical case one
simply drops the term $\frac{15}{4}\hbar ^{2}$ and does not care about
orders of non-commuting factors.

Now, the field $\Psi \left( x^{\mu },\mathbf{y}\right) $ is expanded in
spherical harmonics on S$^{n}$%
\begin{equation}
\Psi \left( x^{\mu },\mathbf{y}\right) =\sum y^{\left( 1-n\right)
/2}f_{l}\left( x^{\mu },y\right) T^{I_{1}\cdots I_{l}}\left( \mathbf{\Omega }%
\right) ,
\end{equation}%
where the rank $l$ traceless tensor $T^{I_{1}\cdots I_{l}}\left( \mathbf{%
\Omega }\right) ,$ constructed by symmetrizing products of $\mathbf{\Omega }%
^{I},$ is the harmonic with angular momentum $l$ on $S^{n}.$ The angular
momentum $\frac{1}{2}L^{IJ}L_{IJ}$ has eigenvalue $\hbar ^{2}l\left(
l+n-2\right) $ on the tensor $T^{I_{1}\cdots I_{l}}\left( \mathbf{\Omega }%
\right) .$ Then $\mathbf{k}^{2}$ reduces to $\mathbf{k}^{2}=k^{2}+\frac{%
l_{n}\left( l_{n}+1\right) \hbar ^{2}}{y^{2}}$ with $l_{n}=l+\left(
n-3\right) /2.$ The $f_{l}\left( x^{\mu },y\right) $ are the radial
wavefunctions that are normalized according to $\int d^{4}xd^{n}\mathbf{y}%
\left\vert \psi \left( x,\mathbf{y}\right) \right\vert ^{2}=1,$ which yields 
$\int_{0}^{\infty }d^{4}xdy\left\vert f_{l}\left( x,y\right) \right\vert
^{2}=1$ after integrating over the orthonormal harmonics $T^{I_{1}\cdots
I_{l}}\left( \mathbf{\Omega }\right) .$ On $f_{l}\left( x^{\mu },y\right) $
the radial momentum $k$ acts as a simple derivative according to $%
kf_{l}\left( y\right) =-i\hbar \partial _{y}f_{l}\left( y\right) .$

For $n=6$ the physical state condition for the field in Eq.(\ref{phys})
reduces to the following differential operator on the radial wavefunction $%
f_{l}\left( x^{\mu },y\right) $ 
\begin{equation}
\,\left[ -\hbar ^{2}y\left( \partial _{\mu }^{2}+\partial _{y}^{2}\right)
y+\left( l+\frac{3}{2}\right) \left( l+\frac{5}{2}\right) \hbar ^{2}\right]
f_{l}=0.
\end{equation}%
This equation can be rewritten as 
\begin{equation}
y^{-5/2}\left[ -\hbar ^{2}\nabla _{AdS_{5}}^{2}+l\left( l+4\right) \hbar ^{2}%
\right] \left[ y^{5/2}f_{l}\right] =0.
\end{equation}%
In the last line $\nabla _{AdS_{5}}^{2}=\left( y^{2}\partial _{\mu
}^{2}+y^{5}\partial _{y}y^{-3}\partial _{y}\right) $ is the Laplacian $\frac{%
1}{\sqrt{-g}}\partial _{\tilde{\mu}}\left( \sqrt{-g}g^{\tilde{\mu}\tilde{\nu}%
}\partial _{\tilde{\nu}}\right) $ for AdS$_{5}$ in our parametrization of
the AdS$_{5}$ metric in Eq.(\ref{metric}). The wavefunction $\phi _{l}\left(
x^{\mu },y\right) =y^{5/2}f_{l}\left( x^{\mu },y\right) $ is normalized
according to the AdS$_{5}$ metric $\int d^{4}xdy\sqrt{-g}\left\vert \phi
_{l}\left( x^{\mu },y\right) \right\vert ^{2}=\int d^{4}xdy\left\vert
f_{l}\left( x^{\mu },y\right) \right\vert ^{2},$ which is consistent with
the above normalization of the radial wavefunction. We see that the last
equation now is precisely the AdS$_{5}\times $S$^{5}$ Laplace equation since 
$l\left( l+4\right) \hbar ^{2}$ is the eigenvalue of the Laplace operator on
S$^{5}$ 
\begin{equation}
\nabla _{\left( AdS_{5}\times S^{5}\right) }^{2}\phi _{l}\left( x^{\mu
},y\right) =0.
\end{equation}%
This is precisely equivalent to the Casimir relations in Eqs.(\ref{c42}-\ref%
{c6}) 
\begin{equation}
\left[ -C_{2}^{\left( 4,2\right) }+C_{2}^{\left( 6,0\right) }\right] \phi
_{l}=0
\end{equation}%
that followed from the gauge invariant Sp$\left( 2,R\right) $ constraints
directly in the 2T formalism in 12 dimensions. Indeed, we already have $%
C_{2}^{\left( 6,0\right) }=\hbar ^{2}l\left( l+4\right) ,$ and if we compute 
$C_{2}^{\left( 4,2\right) }=\frac{1}{2}L_{mn}L^{mn}=\frac{1}{2}L_{\mu \nu
}L^{\mu \nu }-\left( L^{+^{\prime }-^{\prime }}\right) ^{2}-L^{+^{\prime
}\mu }L_{\,\,\,\mu }^{-^{\prime }}-L_{\,\,\,\mu }^{-^{\prime }}L^{+^{\prime
}\mu }$ by inserting an appropriately quantum ordered version of the
non-linear forms in Eqs.(\ref{lmnclass1}-\ref{lmnclass5}), we find precisely 
$C_{2}^{\left( 4,2\right) }=\hbar ^{2}\nabla _{AdS_{5}}^{2}.$

\end{document}